# 3 Xampling: Compressed Sensing of Analog Signals

Moshe Mishali and Yonina C. Eldar

Department of Electrical Engineering, Technion, Haifa 32000, Israel.

This chapter generalizes compressed sensing (CS) to reduced-rate sampling of analog signals. It introduces Xampling, a unified framework for low rate sampling and processing of signals lying in a union of subspaces. Xampling consists of two main blocks: Analog compression that narrows down the input bandwidth prior to sampling with commercial devices followed by a nonlinear algorithm that detects the input subspace prior to conventional signal processing. A variety of analog CS applications are reviewed within the unified Xampling framework including a general filter-bank scheme for sparse shift-invariant spaces, periodic nonuniform sampling and modulated wideband conversion for multiband communications with unknown carrier frequencies, acquisition techniques for finite rate of innovation signals with applications to medical and radar imaging, and random demodulation of sparse harmonic tones. A hardware-oriented viewpoint is advocated throughout, addressing practical constraints and exemplifying hardware realizations where relevant.

## 3.1 Introduction

Analog-to-digital conversion (ADC) technology constantly advances along the route that was delineated in the last century by the celebrated Shannon-Nyquist [1, 2] theorem, essentially requiring the sampling rate to be at least twice the highest frequency in the signal. This basic principle underlies almost all digital signal processing (DSP) applications such as audio, video, radio receivers, wireless communications, radar applications, medical devices, optical systems and more. The ever growing demand for data, as well as advances in radio frequency (RF) technology, have promoted the use of high-bandwidth signals, for which the rates dictated by the Shannon-Nyquist theorem impose demanding challenges on the acquisition hardware and on the subsequent storage and DSP processors. A holy grail of compressed sensing (CS) is to build acquisition devices that exploit signal structure in order to reduce the sampling rate, and subsequent demands on storage and DSP. In such an approach, the actual information contents should dictate the sampling rate, rather than the ambient signal bandwidth. Indeed, CS was motivated in part by the desire to sample wideband signals at rates far below the Shannon-Nyquist rate, while still maintaining the vital information encoded in the underlying signal [3, 4].

At its core, CS is a mathematical framework that studies rate reduction in a discrete setup. A vector $\mathbf{x}$ of length $n$ represents a signal of interest. A measurement vector $\mathbf{y} = \mathbf{A}\mathbf{x}$ is computed using an $m \times n$ matrix $\mathbf{A}$. In a typical CS setup $m \ll n$, so that there are fewer measurements in $\mathbf{y}$ than the ambient dimension of $\mathbf{x}$. Since $\mathbf{A}$ is non-invertible in this setting, recovery must incorporate some prior knowledge on $\mathbf{x}$. The structure that is widely assumed in CS is sparsity, namely that $\mathbf{x}$ has only a few nonzero entries. Convex programming, *e.g.,* $\ell_1$ minimization, and various greedy methods have been shown successful in reconstructing sparse signals $\mathbf{x}$ from short measurement vectors $\mathbf{y}$.

The discrete machinery nicely captures the notion of reduced-rate sampling by the choice $m \ll n$ and affirms robust recovery from incomplete measurements. Nevertheless, since the starting point is of a finite-dimensional vector $\mathbf{x}$, one important aspect is not clearly addressed – how to actually acquire an analog input $x(t)$ at a low rate. In many applications, our interest is to process and represent signals which arrive from the physical domain and are therefore naturally represented as continuous-time functions rather than discrete vectors. A conceptual route to implementing CS in these real-world problems is to first obtain a discrete high-rate representation using standard hardware, and then apply CS to reduce dimensionality. This, however, contradicts the motivation at the heart of CS: reducing acquisition rate as much as possible. Achieving the holy grail of compressive ADCs requires a broader framework which can treat more general signal models including analog signals with various types of structure, as well as practical measurement schemes that can be implemented in hardware. To further gain advantage from the sampling rate decrease, processing speed in the digital domain should also be reduced. Our goal therefore is to develop an end-to-end system, consisting of sampling, processing and reconstruction, where all operations are performed at a low rate, below the Nyquist-rate of the input.

The key to developing low-rate analog sensing methods is relying on structure in the input. Signal processing algorithms have a long history of leveraging structure for various tasks. As an example, MUSIC [5] and ESPRIT [6] are popular techniques for spectrum estimation that exploit signal structure. Model-order selection methods in estimation [7], parametric estimation and parametric feature detection [8] are further examples where structure is heavily exploited. In our context, we are interested in utilizing signal models in order to reduce sampling rate. Classic approaches to sub-Nyquist sampling include carrier demodulation [9], undersampling [10] and nonuniform methods [11–13], which all assume a linear model corresponding to a bandlimited input with predefined frequency support and fixed carrier frequencies. In the spirit of CS, where unknown nonzero locations results in a nonlinear model, we would like to extend the classical treatment to analog inputs with unknown frequency support, as well as more broadly to scenarios that involve nonlinear input structures. The approach we take in this chapter follows the recently proposed Xampling framework [14], which treats a nonlinear model of union of subspaces (UoS). In this structure, originally introduced by Lu an Do [15], the input signal belongs to a single subspace out of multiple, possibly even infinitely many, candidate subspaces. The exact subspace to which the signal belongs is unknown a-priori.

In Section 3.2, we motivate the use of UoS modeling by considering two example sampling problems of analog signals: an RF receiver which intercepts multiple narrowband transmissions, termed multiband communication, but is not provided with their carrier frequencies, and identification of a fading channel which creates echoes of its input at several unknown delays and attenuations. The latter example belongs to a broad model of signals with finite rate of innovation (FRI), discussed in detail in Chapter 4 of this book. FRI models also include other interesting problems in radar and sonar. As we show throughout this chapter, union modeling is a key to savings in acquisition and processing resources.

In Section 3.3, we study a high-level architecture of Xampling systems [14]. The proposed architecture consists of two main functions: lowrate analog to digital conversion (X-ADC) and lowrate digital signal processing (X-DSP). The X-ADC block compresses $x(t)$ in the analog domain, by generating a version of the input that contains all vital information but with relatively lower bandwidth, often substantially below the Nyquist-rate of $x(t)$. The important point is that the chosen analog compression can be efficiently realized with existing hardware components. The compressed version is then sampled at a low rate. X-DSP is responsible for reducing processing rates in the digital domain. To accomplish this goal, the exact signal subspace within the union is detected digitally, using either CS techniques or comparable methods for subspace identification, such as MUSIC [5] or ESPRIT [6]. Identifying the input's subspace allows to execute existing DSP algorithms and interpolation techniques at the low rate of the streaming measurements, that is without going through reconstruction of the Nyquist-rate samples of $x(t)$. Together, when applicable, X-ADC and X-DSP alleviate the Nyquist-rate burden from the entire signal path. Pronounced as CS-Sampling (phonetically /kˈsæmplɪŋ/), the nomenclature Xampling symbolizes the combination between recent developments in CS and the successful machinery of analog sampling theory developed on the past century.

The main body of this chapter is dedicated to study low-rate sampling of various UoS signal models in light of Xampling, capitalizing on the underlying analog model, compressive sensing hardware and digital recovery algorithms. Section 3.4 introduces a framework for sampling sparse shift-invariant (SI) subspaces [16], which extends the classic notion of SI sampling developed for inputs lying in a single subspace [17, 18]. Multiband models [11–13, 19–21] are considered in Section 3.5 with applications to wideband carrier-unaware reception [22] and cognitive radio communication [23]. In particular, this section achieves the X-DSP goal, by considering multiband inputs consisting of a set of digital transmissions whose information bits are recovered and processed at the low rate of the stream-ing samples. Sections 3.6 and 3.7 address FRI signals [24, 25] and sequences of innovation [26], respectively, with applications to pulse stream acquisition and ultrasonic imaging [27, 28]. In radar imaging [29], the Xampling viewpoint not only offers a reduced-rate sampling method, but also allows to increase resolution in target identification and decrease the overall time-bandwidth product of the radar system (when the noise is not too large). Section 3.8 describes sampling strategies that are based on application of CS on discretized analog models, *e.g.,* sampling a sparse sum of harmonic tones [30] and works on quantized CS radar [31–33].

Besides reviewing sampling strategies, we provide some insights into analog sensing. In Section 3.5, we use the context of multiband sampling to exemplify a full development cycle of analog CS systems, from theory to hardware. The cycle begins with a nonuniform method [19] that is derived from the sparse-SI framework. Analyzing this approach in a more practical perspective, reveals that nonuniform acquisition requires ADC devices with Nyquist-rate frontend since they are connected directly to the wideband input. We next review the hardware-oriented design of the modulated wideband converter (MWC) [20, 22], which incorporates RF preprocessing to compress the wideband input, so that actual sampling is carried out by commercial lowrate and low bandwidth ADC devices. To complete the cycle, we take a glimpse at circuit challenges and solutions as reported in the design of an MWC hardware prototype [22]. The MWC appears to be the first reported wideband technology borrowing CS ideas with provable hardware that samples and processes wideband signals at a rate that is directly proportional to the actual bandwidth occupation and not the highest frequency (280 MHz sampling of 2 GHz Nyquist-rate inputs in [22]).

Xampling advocates use of traditional tools from sampling theory for modeling analog signals, according to which a continuous-time signal $x(t)$ is determined by a countable sequence $c[n]$ of numbers, *e.g.,* a bandlimited input $x(t)$ and its equally-spaced pointwise values $c[n] = x(nT)$. The UoS approach is instrumental in capturing similar infinite structures by taking to infinity either the dimensions of the individual subspaces, the number of subspaces in the union or both. In Section 3.8 we review alternative analog CS methods which treat continuous signals that are determined by a finite set of parameters. This approach was taken, for example, in the development of the random demodulator (RD) [30] and works on quantized CS radar [31–33]. Whilst effective in the finite scenarios for which they were developed, the application of these methods to general analog models (which possess a countable representation) can lead to performance degradation. We exemplify differences when comparing hardware and software complexities of the RD and MWC systems. Visualizing radar performance of quantized [33] vs. analog [29] approaches further demonstrates the possible differences. Based on the insights gained throughout this chapter, several operative conclusions are suggested in Section 3.9 for extending CS to general analog signals.

## 3.2   From Subspaces to Unions

The traditional paradigm in sampling theory assumes that $x(t)$ lies in a single subspace. Bandlimited sampling is undoubtedly the most studied example. Subspace modeling is quite powerful, as it allows perfect recovery of the signal from its linear and nonlinear samples under very broad conditions [17,18,34–36]. Furthermore, recovery can be achieved by digital and analog filtering. This is a very appealing feature of the subspace model, which generalizes the Shannon-Nyquist theorem to a broader set of input classes.

Despite the simplicity and intuitive appeal of subspace modeling, in modern applications many signals are characterized by parameters which are not necessarily known to the sampler. As we will show now via several examples, we can often still describe the signal by a subspace model. However, in order to include all possible parameter choices, the subspace has to have large dimension with enough degrees of freedom to capture the uncertainty, leading to extremely high sampling rates. The examples below build the motivation for lowrate sampling solutions which we discuss in the rest of this chapter.

Consider first the scenario of a multiband input $x(t)$, which has sparse spectra, such that its continuous-time Fourier transform (CTFT) $X(f)$ is supported on $N$ frequency intervals, or bands, with individual widths not exceeding $B$ Hz. Figure 3.1 illustrates a typical multiband spectra. When the band positions are known and fixed, the signal model is linear, since the CTFT of any combination of two inputs is supported on the same frequency bands. This scenario is typical in communication, when a receiver intercepts several RF transmissions, each modulated on a different high carrier frequency $f_i$. Knowing the band positions, or the carriers $f_i$, allows the receiver to demodulate a transmission of interest to baseband, that is to shift the contents from the relevant RF band to the origin. Several demodulation topologies are reviewed in [37]. Subsequent sampling and processing are carried out at the low rate corresponding to the individual band of interest. When the input consists of a single transmission, an alternative approach to shift contents to baseband is by uniform undersampling at a properly chosen sub-Nyquist rate [10]. Nonuniform sampling methods that can treat more than a single transmission were developed in [12,13], under the assumption that the digital recovery algorithm is provided with knowledge of the spectral support.

When the carrier frequencies $f_i$ are unknown, we are interested in the set of all possible multiband signals that occupy up to $NB$ Hz of the spectrum. In this scenario, the transmissions can lie anywhere below $f_{\max}$. At first sight, it may seem that sampling at the Nyquist rate

$$f_{\mathrm{NYQ}} = 2f_{\max}, \tag{3.1}$$

is necessary, since every frequency interval below $f_{\max}$ appears in the support of some multiband $x(t)$. On the other hand, since each specific $x(t)$ in this model fills only a portion of the Nyquist range (only $NB$ Hz), we intuitively expect to be able to reduce the sampling rate below $f_{\text{NYQ}}$. Standard demodulation cannot be used since $f_i$ are unknown, which makes this sampling problem challenging.

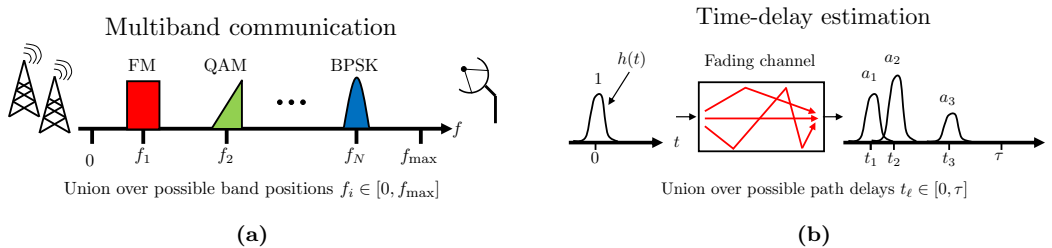

**Figure 3.1:** Example applications of UoS modeling.

Another interesting application is estimation of time delays from observation of a signal of the following form

$$x(t) = \sum_{\ell=1}^{L} a_\ell\, h(t - t_\ell), \quad t \in [0, \tau]. \tag{3.2}$$

For fixed time delays $t_\ell$, (3.2) defines a linear space of inputs with $L$ degrees of freedom, one per each amplitude $a_\ell$. In this case, $L$ samples of $x(t)$ can be used to reconstruct the input $x(t)$. In practice, however, there are many interesting situations with unknown $t_\ell$. Inputs of this type belong to the broader family of FRI signals [24, 25], and are treated in detail in Chapter 4 of this book. For example, when a communication channel introduces multipath fading, the transmitter can assist the receiver in channel identification by sending a short probing pulse $h(t)$. Since the receiver knows the shape of $h(t)$, it can resolve the delays $t_\ell$ and use this information to decode the following information messages. Another example is radar, where the delays $t_\ell$ correspond to target locations, while the amplitudes $a_\ell$ encode Doppler shifts indicating target speeds. Medical imaging techniques, *e.g.*, ultrasound, use signals of the form (3.2) to probe density changes in human tissues as a vital tool in medical diagnosis. Underwater acoustics also conform with (3.2). Since in all these applications, the pulse $h(t)$ is short in time, sampling $x(t)$ according to its Nyquist bandwidth, which is effectively that of $h(t)$, results in unnecessary large sampling rates. In contrast, it follows intuitively from (3.2), that only $2L$ unknowns determine $x(t)$, namely $t_\ell, a_\ell, 1 \leq \ell \leq L$. Since with unknown delays, (3.2) describes a nonlinear model, subspace modeling cannot achieve the optimal sampling rate of $2L/\tau$, which in all the above applications can be substantially lower than Nyquist.

The example applications above motivate the need for signal modeling that is more sophisticated than the conventional single subspace approach. In order to capture real-world scenarios within a convenient mathematical formulation without unnecessarily increasing the rate, we introduce in the next section the Xampling framework which treats UoS signal classes and is applicable to many interesting applications. Using the Xampling framework, we will analyze sampling strategies for several union models in detail, and show that although sampling can still be obtained by linear filtering, recovery becomes more involved and requires nonlinear algorithms, following the spirit of CS.

## 3.3  Xampling

In this section, we introduce Xampling – our proposed framework for acquisition and digital processing of UoS signal models [14].

### 3.3.1   Union of Subspaces

As motivated earlier, the key to reduced-rate sampling of analog signals is based on UoS modeling of the input set. The concept of allowing more than a single input subspace was first suggested by Lu and Do in [15]. We denote by $x(t)$ an analog signal in the Hilbert space $\mathcal{H} = L_2(\mathbb{R})$, which lies in a parameterized family of subspaces

$$x(t) \in \mathcal{U} \triangleq \bigcup_{\lambda \in \Lambda} \mathcal{A}_\lambda, \tag{3.3}$$

where $\Lambda$ is an index set, and each individual $\mathcal{A}_\lambda$ is a subspace of $\mathcal{H}$. The key property of the UoS model (3.3) is that the input $x(t)$ resides within $\mathcal{A}_{\lambda^*}$ for some $\lambda^* \in \Lambda$, but a-priori, the exact subspace index $\lambda^*$ is unknown. For example, multiband signals with unknown carriers $f_i$ can be described by (3.3), where each $\mathcal{A}_\lambda$ corresponds to signals with specific carrier positions and the union is taken over all possible $f_i \in [0, f_{\max}]$. Pulses with unknown time-delays of the form (3.2) also obey UoS modeling, where each $\mathcal{A}_\lambda$ is an $L$ dimensional subspace that captures the coefficients $a_\ell$, whereas the union over all possible delays $t_\ell \in [0, \tau]$ provides an efficient way to group these subspaces to a single set $\mathcal{U}$.

UoS modeling enables treating $x(t)$ directly in its analog formulation. This approach is fundamentally different than previous attempts to treat similar problems, which rely on discretization of the analog input to finite representations. Namely, models in which both cardinalities, $\Lambda$ and each $\mathcal{A}_\lambda$, are finite. Standard CS which treats vectors in $\mathbb{R}^n$ having at most $k$ nonzeros is a special case of a finite representation. Each individual subspace has dimensions $k$, defined by the locations of the nonzeros, and the union is over $\binom{n}{k}$ possibilities of choosing the nonzero locations. In Section 3.8, we discuss in detail the difference between union modeling and discretization. As we show, the major consequences of imposing a finite representation on an analog signal that does not inherently conform to a finite model are twofold: model sensitivity and high computational loads. Therefore, the main core of this chapter focuses on the theory and applications developed for general UoS modeling (3.3). We note that there are examples of continuous-time signals that naturally possess finite representations. One such example are trigonometric polynomials. However, our interest here is in signals of the form described in Section 3.2, that do not readily admit a finite representation.

The union (3.3) over all possible signal locations forms a nonlinear signal set $\mathcal{U}$, where its nonlinearity refers to the fact that the sum (or any linear combination) of $x_1(t), x_2(t) \in \mathcal{U}$ does not lie in $\mathcal{U}$, in general. Consequently, $\mathcal{U}$ is a true subset of the linear affine space

$$\Sigma = \left\{ x(t) = \sum_{\lambda \in \Lambda} \alpha_\lambda x_\lambda(t) \,:\, \alpha_\lambda \in \mathbb{R},\, x_\lambda(t) \in \mathcal{A}_\lambda \right\}, \quad (3.4)$$

which we refer to as the Nyquist subspace of $\mathcal{U}$. Since every $x(t) \in \mathcal{U}$ also belongs to $\Sigma$, one can in principle apply conventional sampling strategies with respect to the single subspace $\Sigma$ [18]. However, this technically-correct approach often leads to practically-infeasible sampling systems with a tremendous waste of expensive hardware and software resources. For example, in multiband sampling, $\Sigma$ is the $f_{\max}$-bandlimited space, for which no rate reduction is possible. Similarly, in time-delay estimation problems, $\Sigma$ has the high bandwidth of $h(t)$, and again no rate reduction can be achieved.

We define the sampling problem for the union set (3.3) as the design of a system that provides:

1. **ADC:** an acquisition operator which converts the analog input $x(t) \in \mathcal{U}$ to a sequence $y[n]$ of measurements,
2. **DSP:** a toolbox of processing algorithms, which uses $y[n]$ to perform classic tasks, *e.g.,* estimation, detection, data retrieval etc., and
3. **DAC:** a method for reconstructing $x(t)$ from the samples $y[n]$.

In order to exclude from consideration inefficient solutions, such as those treating the Nyquist subspace $\Sigma$ and not exploiting the union structure, we adopt as a general design constraint that the above goals should be accomplished with minimum use of resources. Minimizing the sampling rate, for example, excludes inefficient Nyquist-rate solutions and tunnel potential approaches to wisely incorporate the union structure to stand this resource constraint. For reference, this requirement is outlined as

$$\textbf{ADC} + \textbf{DSP} + \textbf{DAC} \rightarrow \textbf{minimum use of resources}. \quad (3.5)$$

In practice, besides constraining the sampling rate, (3.6) translates to the minimization of several other resources of interest, including the number of devices in the acquisition stage, design complexity, processing speed, memory requirements, power dissipation, system cost, and more. As we shall see via examples in the sequel, the challenge posed in (3.6) is to treat a union model at an overall complexity (of hardware and software) that is comparable with a system which knows the exact $\mathcal{A}_{\lambda^*}$.

In order to exclude from consideration inefficient solutions, such as those treating the Nyquist subspace $\Sigma$ and not exploiting the union structure, we adopt as a general design constraint that the above goals should be accomplished with minimum use of resources. Minimizing the sampling rate, for example, excludes inefficient Nyquist-rate solutions and tunnel potential approaches to wisely incorporate the union structure to stand this resource constraint. For reference, this requirement is outlined as

$$\textbf{ADC} + \textbf{DSP} + \textbf{DAC} \to \textbf{minimum use of resources}. \qquad (3.6)$$

In practice, besides constraining the sampling rate, (3.6) translates to the minimization of several other resources of interest, including the number of devices in the acquisition stage, design complexity, processing speed, memory requirements, power dissipation, system cost, and more. As we shall see via examples in the sequel, the challenge posed in (3.6) is to treat a union model at an overall complexity (of hardware and software) that is comparable with a system which knows the exact $\mathcal{A}_{\lambda^*}$.

In essence, the UoS model follows the spirit of classic sampling theory by assuming that $x(t)$ belongs to a single underlying subspace $\mathcal{A}_{\lambda^*}$. However, in contrast to the traditional paradigm, the union setting permits uncertainty in the exact signal subspace, opening the door to interesting sampling problems. The challenge posed in (3.6) is to treat the uncertainty of the union model at an overall complexity (of hardware and software) that is comparable with a system which knows the exact $\mathcal{A}_{\lambda^*}$. In Section 3.5, we describe strategies which acquire and process signals from the multiband union at a low rate, proportional to $NB$. Sections 3.6 and 3.7 describe variants of FRI unions, including (3.2), and their lowrate sampling solutions, which approach the rate of innovation $2L/\tau$. A line of other UoS applications that are described throughout this chapter exhibit similar rationale – the sampling rate is reduced by exploiting the fact that the input belongs to a single subspace $\mathcal{A}_{\lambda^*}$, even though the exact subspace index $\lambda^*$ is unknown. The next subsection proposes a systematic architecture for the design of sampling systems for UoS signal classes. As we show in the ensuing sections, this architecture unifies a variety of sampling strategies developed for different instances of UoS models.

### 3.3.2 Architecture

The Xampling system we propose has the high-level architecture presented in Fig. 3.2 [14]. The first two blocks, termed X-ADC, perform the conversion of $x(t)$ to digital. An operator $P$ compresses the high-bandwidth input $x(t)$ into a signal with lower bandwidth, effectively capturing the entire union $\mathcal{U}$ by a subspace $\mathcal{S}$ with substantially lower sampling requirements. A commercial ADC device then takes pointwise samples of the compressed signal, resulting in the sequence of samples $y[n]$. The role of $P$ in Xampling is to narrow down the analog bandwidth, so that lowrate ADC devices can subsequently be used. As in digital compression, the goal is to capture all vital information of the input in the compressed version, though here this functionality is achieved by hardware rather than software. The design of $P$ therefore needs to wisely exploit the union structure, in order not to lose any essential information while reducing the bandwidth.

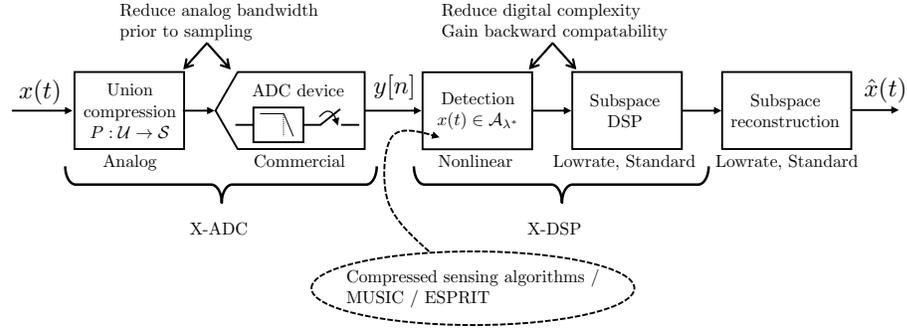

**Figure 3.2:** Xampling – A pragmatic framework for signal acquisition and processing in union of subspaces (taken from [14]).

In the digital domain, Xampling consists of three computational blocks. A nonlinear step detects the signal subspace $\mathcal{A}_{\lambda^*}$ from the lowrate samples. CS algorithms, *e.g.,* those described in the relevant chapters of this book, as well as comparable methods for subspace identification, *e.g.,* MUSIC [5] or ESPRIT [6], can be used for that purpose. Once the index $\lambda^*$ is determined, we gain backward compatibility, meaning standard DSP methods apply and commercial DAC devices can be used for signal reconstruction. The combination of nonlinear detection and standard DSP is referred to as X-DSP. As we demonstrate, besides backward compatibility, the nonlinear detection decreases computational loads, since the subsequent DSP and DAC stages need to treat only the single subspace $\mathcal{A}_{\lambda^*}$, complying with (3.6). The important point is that the detection stage can be performed efficiently at the low acquisition rate, without requiring Nyquist-rate processing.

Xampling is a generic template architecture. It does not specify the exact acquisition operator $P$ or nonlinear detection method to be used. These are application-dependant functions. Our goal in introducing Xampling is to propose a high-level system architecture and a basic set of guidelines:

1. an analog pre-processing unit to compress the input bandwidth,
2. commercial lowrate ADC devices for actual acquisition at a low rate,
3. subspace detection in software, and
4. standard DSP and DAC methods.

The Xampling framework is developed in [14] based on two basic assumptions:

(**A1**) DSP is the main purpose of signal acquisition, and
(**A2**) The ADC device has limited bandwidth.

The DSP assumption (**A1**) highlights the ultimate use of many sampling systems – substituting analog processing by modern software algorithms. DSP is perhaps the most profound reason for signal acquisition: Hardware development can rarely compete with the convenience and flexibilities that software environments provide. In many applications, therefore, DSP is what essentially motivates the ADC and decreasing processing speeds can sometimes be an important requirement, regardless of whether the sampling rate is reduced as well. In particular, the digital flow proposed in Fig. 3.2 is beneficial even when a high ADC rate is acceptable. In this case, $x(t)$ can be acquired directly without narrowing down its bandwidth prior to ADC, but we would still like to reduce computational loads and storage requirements in the digital domain. This can be accomplished by imitating rate reduction in software, detecting the signal subspace and processing at the actual information bandwidth. The compounded usage of both X-ADC and X-DSP is for mainstream applications, where reducing the rate of both signal acquisition and processing is of interest.

Assumption (**A2**) basically says that we expect the conversion device to have limited front-end bandwidth. The X-ADC can be realized on a circuit board, chip design, optical system or other appropriate hardware. In all these platforms, the front-end has certain bandwidth limitations which obey (**A2**), thereby motivating the use of a preceding analog compression step $P$ in order to capture all vital information within a narrow range of frequencies that the acquisition device can handle. Section 3.5 elaborates on this property.

Considering the architecture of Fig. 3.2 in conjunction with requirement (3.6) reveals an interesting aspect of Xampling. In standard CS, most of the system complexity concentrates in digital reconstruction, since sensing is as simple as applying $\mathbf{y} = \mathbf{Ax}$. In Xampling, we attempt to balance between analog and digital complexities. As discussed in Section 3.8, a properly chosen analog pre-processing operator $P$ can lead to substantial savings in digital complexities and vice versa.

We next describe sampling solutions for UoS models according to the Xampling paradigm. In general, when treating unions of analog signals, there are three main cases to consider:

- finite unions of infinite dimensional spaces;
- infinite unions of finite dimensional spaces;
- infinite unions of infinite dimensional spaces.

In each one of the three settings above there is an element that can take on infinite values, which is a result of the fact that we are considering general analog signals: either the underlying subspaces $\mathcal{A}_\lambda$ are infinite-dimensional, or the number of subspaces $|\Lambda|$ is infinite. In the next sections, we present general theory and results behind each of these cases, and focus in additional detail on a representative example application for each class. Sections 3.4 and 3.5 cover the first scenario, introducing the sparse-SI framework and reviewing multiband sampling strategies, respectively. Sections 3.6 and 3.7 discuss variants of inno-vation rate sampling and cover the other two cases. Methods that are based on completely finite unions, when both $|\Lambda|$ and $\mathcal{A}_\lambda$ are finite, are discussed in Section 3.8. While surveying these different cases, we will attempt to shed light into pragmatic considerations that underlie Xampling, and hint on possible routes to promote these compressive methods to actual hardware realizations.

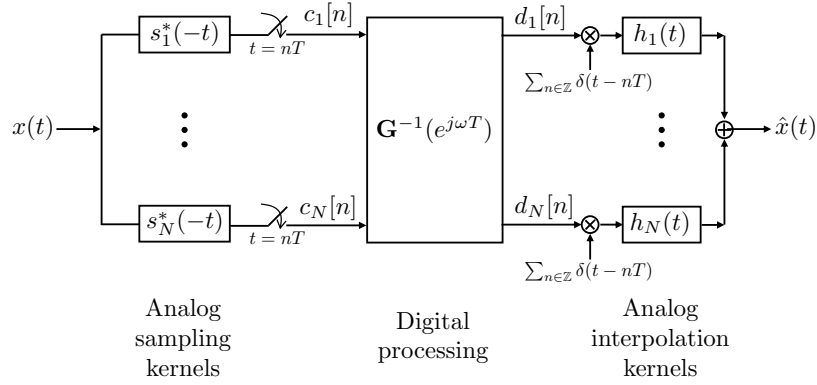

**Figure 3.3:** Sampling and reconstruction in shift-invariant spaces [16, 34] (taken from [16]).

## 3.4   Sparse Shift-Invariant Framework

### 3.4.1   Sampling in Shift-Invariant Subspaces

We first briefly introduce the notion of sampling in SI subspaces, which plays a key role in the development of standard (subspace) sampling theory [17, 18]. We then discuss how to incorporate the union structure into SI settings.

SI signals are characterized by a set of generators $\{h_\ell(t), 1 \leq \ell \leq N\}$ where in principle $N$ can be finite or infinite (as is the case in Gabor or wavelet expansions of $L_2$). Here we focus on the case in which $N$ is finite. Any signal in such an SI space can be written as

$$x(t) = \sum_{\ell=1}^{N} \sum_{n \in \mathbb{Z}} d_\ell[n] h_\ell(t - nT), \qquad (3.7)$$

for some set of sequences $\{d_\ell[n] \in \ell_2, 1 \leq \ell \leq N\}$ and period $T$. This model encompasses many signals used in communication and signal processing including bandlimited functions, splines [38], multiband signals (with known carrier positions) [11, 12] and pulse amplitude modulation signals.

The subspace of signals described by (3.7) has infinite dimensions, since every signal is associated with infinitely many coefficients $\{d_\ell[n], 1 \leq \ell \leq N\}$. Any such signal can be recovered from samples at a rate of $N/T$; one possible sampling paradigm at the minimal rate is given in Fig. 3.3 [16, 34].

Here $x(t)$ is filtered with a bank of $N$ filters, each with impulse response $s_\ell(t)$ which can be almost arbitrary. The outputs are uniformly sampled with period $T$, resulting in the sample sequences $c_\ell[n]$. Denote by $\mathbf{c}(\omega)$ a vector collecting the frequency responses of $c_\ell[n]$, $1 \leq \ell \leq N$, and similarly $\mathbf{d}(\omega)$ for the frequency responses of $d_\ell[n]$, $1 \leq \ell \leq N$. Then, it can be shown that [16]

$$\mathbf{c}(\omega) = \mathbf{G}(e^{j\omega T})\mathbf{d}(\omega), \tag{3.8}$$

where $\mathbf{G}(e^{j\omega T})$ is an $N \times N$ matrix, with entries

$$\left[\mathbf{G}(e^{j\omega T})\right]_{i\ell} = \frac{1}{T}\sum_{k\in\mathbb{Z}} S_i^*\left(\frac{\omega}{T} - \frac{2\pi}{T}k\right) H_\ell^*\left(\frac{\omega}{T} - \frac{2\pi}{T}k\right). \tag{3.9}$$

The notations $S_i(\omega), H_\ell(\omega)$ stand for the CTFT of $s_i(t), h_\ell(t)$, respectively. To allow recovery, the condition on the sampling filters $s_i(t)$ is that (3.9) results in an invertible frequency response $\mathbf{G}(e^{j\omega T})$. The signal is then recovered by processing the samples with a filter bank with frequency response $\mathbf{G}^{-1}(e^{j\omega T})$. In this way, we invert (3.8) and obtain the vectors

$$\mathbf{d}(\omega) = \mathbf{G}^{-1}(e^{j\omega T})\mathbf{c}(\omega). \tag{3.10}$$

Each output sequence $d_\ell[n]$ is then modulated by a periodic impulse train $\sum_{n\in\mathbb{Z}} \delta(t - nT)$ with period $T$, followed by filtering with the corresponding analog filter $h_\ell(t)$. In practice, interpolation with finitely many samples gives sufficiently accurate reconstruction, provided that $h_\ell(t)$ decay fast enough [39], similar to finite interpolation in the Shannon-Nyquist theorem.

### 3.4.2 Sparse Union of SI Subspaces

In order to incorporate further structure into the generic SI model (3.7), we treat signals of the form (3.7) involving a small number $K$ of generators, chosen from a finite set $\Lambda$ of $N$ generators. Specifically, we consider the input model

$$x(t) = \sum_{|\ell|=K} \sum_{n\in\mathbb{Z}} d_\ell[n] h_\ell(t - nT), \tag{3.11}$$

where $|\ell| = K$ means a sum over at most $K$ elements. If the $K$ active generators are known, then according to Fig. 3.3 it suffices to sample at a rate of $K/T$ corresponding to uniform samples with period $T$ at the output of $K$ appropriate filters. A more difficult question is whether the rate can be reduced if we know that only $K$ of the generators are active, but do not know in advance which ones. In terms of (3.11) this means that only $K$ of the sequences $d_\ell[n]$ have nonzero energy. Consequently, for each value $n$, $\|\mathbf{d}[n]\|_0 \leq K$, where $\mathbf{d}[n] = [d_1[n], \cdots, d_N[n]]^T$ collects the unknown generator coefficients for time instance $n$.

For this model, it is possible to reduce the sampling rate to as low as $2K/T$ [16] as follows. We target a compressive sampling system that produces a vector of

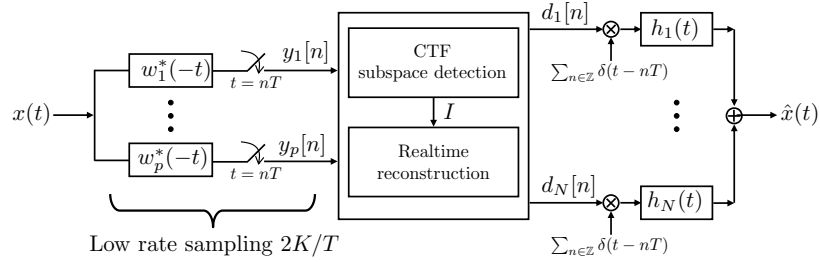

**Figure 3.4:** Compressive sensing acquisition for sparse union of shift-invariant subspaces (taken from [16]).

lowrate samples $\mathbf{y}[n] = [y_1[n], \cdots, y_p[n]]^T$ at $t = nT$ which satisfies a relation

$$\mathbf{y}[n] = \mathbf{A}\mathbf{d}[n], \quad \|\mathbf{d}[n]\|_0 \leq K, \tag{3.12}$$

with a sensing matrix $\mathbf{A}$ that allows recovery of sparse vectors. The choice $p < N$ reduces the sampling rate below Nyquist. In principle, a parameterized family of underdetermined systems, by the time index $n$ in the case of (3.12), can be treated by applying CS recovery algorithms independently for each $n$. A more robust and efficient technique which exploits the joint sparsity over $n$ is described in the next section. The question is therefore how to design a sampling scheme which would boil down to a relation such as (3.12) in the digital domain. Figure 3.4 provides a system for obtaining $\mathbf{y}[n]$, where the following theorem gives the expression for its sampling filters $w_\ell(t)$ [16].

**Theorem 3.1.** *Let $s_\ell(t)$ be a set of $N$ filters and $\mathbf{G}(e^{j\omega T})$ the response matrix defined in (3.9) (so that $s_\ell(t)$ can be used in the Nyquist-rate scheme of Fig. 3.3), and let $\mathbf{A}$ be a given $p \times N$ sensing matrix. Sampling $x(t)$ with a bank of filters $w_\ell(t), 1 \leq \ell \leq p$ defined by*

$$\mathbf{w}(\omega) = \mathbf{A}^* \mathbf{G}^{-*}(e^{j\omega T})\mathbf{s}(\omega), \tag{3.13}$$

*gives a set of compressed measurements $y_\ell[n], 1 \leq \ell \leq p$ that satisfies (3.12). In (3.13), the vectors $\mathbf{w}(\omega), \mathbf{s}(\omega)$ have $\ell$th elements $W_\ell(\omega), S_\ell(\omega)$, denoting CTFTs of the corresponding filters, and $(\cdot)^{-*}$ denotes the conjugate of the inverse.*

The filters $w_\ell(t)$ of Fig. 3.4 form an analog compression operator $P$ as suggested in the X-ADC architecture. The sampling rate is effectively reduced by taking linear combinations of the outputs $c_\ell[n]$ of the Nyquist scheme of Fig. 3.3, with combination coefficients defined by the sensing matrix $\mathbf{A}$. This structure is revealed by examining (3.13) – sampling by $w_\ell[n]$ is tantamount to filtering $x(t)$ by $s_\ell(t)$, applying $\mathbf{G}^{-1}(e^{j\omega T})$ to obtain the sparse set of sequences $d_\ell[n]$, and then combining these sequences by an underdetermined matrix $\mathbf{A}$. A more general result of [16] enables further flexibility in choosing the sampling filters by letting $\mathbf{w}(\omega) = \mathbf{P}^*(e^{j\omega T})\mathbf{A}^* \mathbf{G}^*(e^{j\omega T})\mathbf{s}(\omega)$, for some arbitrary invertible $p \times p$ matrix $\mathbf{P}^*(e^{j\omega T})$. In this case, (3.12) holds with respect to sequences obtained by post-processing the compressive measurements $y_\ell[n]$ by $\mathbf{P}^{-1}(e^{j\omega T})$.

The sparse-SI model (3.11) can be generalized to a sparse sum of arbitrary subspaces, where each subspace $\mathcal{A}_\lambda$ of the union (3.3) consists of a direct sum of $K$ low-dimensional subspaces [40]

$$\mathcal{A}_\lambda = \bigoplus_{|j|=K} \mathcal{V}_j. \tag{3.14}$$

Here $\{\mathcal{V}_j, 1 \leq j \leq N\}$ are a given set of subspaces with dimensions $\dim(\mathcal{V}_j) = v_j$, and as before $|j| = K$ denotes a sum over $K$ indices. Thus, each subspace $\mathcal{A}_\lambda$ corresponds to a different choice of $K$ subspaces $\mathcal{V}_j$ that comprise the sum. The sparse-SI model is a special case of (3.14), in which each $\mathcal{V}_j$ is an SI subspace with a single shift kernel $h_j(t)$. In [40], sampling and reconstruction algorithms are developed for the case of finite $\Lambda$ and finite-dimensional $\mathcal{A}_\lambda$. The approach utilizes the notion of set transforms to cast the sampling problem into an underdetermined system with an unknown block-sparse solution, which is found via a polynomial-time mixed-norm optimization program. Block-sparsity is studied in more detail in [40–43].

### 3.4.3  Infinite Measurement Model and Continuous to Finite

In the sparse-SI framework, the acquisition scheme is mapped into the system (3.12). Reconstruction of $x(t)$ therefore depends on our ability to resolve $d_\ell[n]$ from this underdetermined system. More generally, we are interested in solving a parameterized underdetermined linear system with sensing matrix dimensions $p \times N, p < N$

$$\mathbf{y}(\theta) = \mathbf{A}\mathbf{x}(\theta), \quad \theta \in \Theta, \tag{3.15}$$

where $\Theta$ is a set whose cardinality can be infinite. In particular, $\Theta$ may be uncountable, such as the frequencies $\omega \in [-\pi, \pi)$ of (3.13), or countable as in (3.12). The system (3.15) is referred to as an infinite measurement vector (IMV) model with sparsity $K$, if the vectors $\mathbf{x}(\Theta) = \{\mathbf{x}(\theta)\}$ share a joint sparsity pattern [44]. That is, the non-zero elements are supported within a fixed location set $I$ of size $K$.

The IMV model includes as a special case standard CS, when taking $\Theta = \{\theta^*\}$ to be a single element set. It also includes the case of a finite set $\Theta$, termed multiple measurement vectors (MMV) in the CS literature [44–49]. In the finite cases it is easy to see that if $\sigma(\mathbf{A}) \geq 2k$, where $\sigma(\mathbf{A}) = \mathrm{spark}(\mathbf{A}) - 1$ is the Kruskal-rank of $\mathbf{A}$, then $\mathbf{x}(\Theta)$ is the unique $K$-sparse solution of (3.15) [47]. A simple necessary and sufficient condition in terms of $\mathrm{rank}(\mathbf{y}(\Theta))$ is derived in [50], which improves upon earlier (sufficient only) conditions in [47]. Similar conditions hold for a jointly $K$-sparse IMV system [44].

The major difficulty with the IMV model is how to recover the solution set $\mathbf{x}(\Theta)$ from the infinitely many equations (3.15). One strategy is to solve (3.15) independently for each $\theta$. However, this strategy may be computationally intensive in practice, since it would require to execute a CS solver for each individual $\theta$; for example, in the context of (3.12), this amounts to solving a sparse recovery problem for each time instance $n$. A more efficient strategy exploits the fact that $\mathbf{x}(\Theta)$ are jointly sparse, so that the index set

$$I = \{l \,:\, x_l(\theta) \neq 0\}, \tag{3.16}$$

is independent of $\theta$. Therefore, $I$ can be estimated from several instances of $\mathbf{y}(\Theta)$, which increases the robustness of the estimate. Once $I$ is found, recovery of the entire set $\mathbf{x}(\Theta)$ is straightforward. To see this, note that using $I$, (3.15) can be written as

$$\mathbf{y}(\theta) = \mathbf{A}_I \mathbf{x}_I(\theta), \quad \theta \in \Theta, \tag{3.17}$$

where $\mathbf{A}_I$ denotes the matrix containing the columns of $\mathbf{A}$ whose indices belong to $I$, and $\mathbf{x}_I(\theta)$ is the vector consisting of entries of $\mathbf{x}(\theta)$ in locations $I$. Since $\mathbf{x}(\Theta)$ is $K$-sparse, $|I| \leq K$. Therefore, the columns of $\mathbf{A}_I$ are linearly independent (because $\sigma(\mathbf{A}) \geq 2K$), implying that $\mathbf{A}_I^\dagger \mathbf{A}_I = \mathbf{I}$, where $\mathbf{A}_I^\dagger = \left(\mathbf{A}_I^H \mathbf{A}_I\right)^{-1} \mathbf{A}_I^H$ is the pseudo-inverse of $\mathbf{A}_I$ and $(\cdot)^H$ denotes the Hermitian conjugate. Multiplying (3.17) by $\mathbf{A}_I^\dagger$ on the left gives

$$\mathbf{x}_I(\theta) = \mathbf{A}_I^\dagger \mathbf{y}(\theta), \quad \theta \in \Theta. \tag{3.18}$$

The components in $\mathbf{x}(\theta)$ not supported on $S$ are all zero. In contrast to applying a CS solver for each $\theta$, (3.18) requires only one matrix-vector multiplication per $\mathbf{y}(\theta)$, typically requiring far fewer computations.

It remains to determine $I$ efficiently. In [44] it was shown that $I$ can be found exactly by solving a finite MMV. The steps used to formulate this MMV are grouped under a block referred to as continuous-to-finite (CTF). The essential idea is that every finite collection of vectors spanning the subspace span($\mathbf{y}(\Theta)$) contains sufficient information to recover $I$, as incorporated in the following theorem [44]:

**Theorem 3.2.** *Suppose that $\sigma(\mathbf{A}) \geq 2K$, and let $\mathbf{V}$ be a matrix with column span equal to* span($\mathbf{y}(\Theta)$). *Then, the linear system*

$$\mathbf{V} = \mathbf{A}\mathbf{U} \tag{3.19}$$

*has a unique $K$-sparse solution $\mathbf{U}$ whose support is equal $I$.*

The advantage of Theorem 3.2 is that it allows to avoid the infinite structure of (3.15) and instead find the finite set $I$ by solving a single MMV system of the form (3.19).

For example, in the sparse SI model, such a frame can be constructed by

$$\mathbf{Q} = \sum_n \mathbf{y}[n]\mathbf{y}^H[n], \tag{3.20}$$

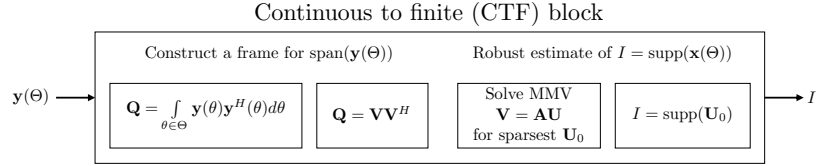

**Figure 3.5:** The fundamental stages for the recovery of the non-zero location set $I$ in an IMV model using only one finite-dimensional program (taken from [44]).

where typically $2K$ snapshots $\mathbf{y}[n]$ are sufficient [20]. Optionally, $\mathbf{Q}$ is decomposed to another frame $\mathbf{V}$, such that $\mathbf{Q} = \mathbf{V}\mathbf{V}^H$, allowing removal of the noise space [20]. Applying the CTF in this setting provides a robust estimate of $I = \mathrm{supp}(d_\ell[n])$, namely the indices of the active generators that comprise $x(t)$. This is essentially the subspace detection part of X-DSP, where the joint support set $I$ determines the signal subspace $\mathcal{A}_{\lambda^*}$. The crux of the CTF now becomes apparent – the indices of the nonidentically-zero rows of the matrix $\mathbf{U}_0$ that solves the finite underdetermined system (3.19) coincide with the index set $I = \mathrm{supp}(d_\ell[n])$ that is associated with the continuous signal $x(t)$ [44]. Once $I$ is found, (3.12) can be inverted on the column subset $I$ by (3.18), where the time index $n$ takes the role of $\theta$. Reconstruction from that point on is carried out in real time; one matrix-vector multiplication (3.18) per incoming vector of samples $\mathbf{y}[n]$ recovers $\mathbf{d}_I[n]$, denoting the entries of $\mathbf{d}[n]$ indicated by $I$.

Figure 3.5 summarizes the CTF steps for identifying the nonzero location set of an IMV system. In the figure, the summation (3.20) is formulated as integration over $\theta \in \Theta$ for the general IMV setting (3.15). The additional requirement of Theorem 3.2 is to construct a frame matrix $\mathbf{V}$ having column span equal to $\mathrm{span}(\mathbf{y}(\Theta))$, which, in practice, is computed efficiently from the samples.

The mapping of Fig. 3.4 to an IMV system (3.12) and the CTF recovery create a nice connection to results of standard CS. The number of branches $p$ is the number of rows in $\mathbf{A}$, and the choice of sampling filters $w_\ell(t)$ translate to its entries via Theorem 3.1. Since recovery boils down to solving an MMV system with sensing matrix $\mathbf{A}$, we should design the hardware so that the resulting matrix $\mathbf{A}$ in (3.13) has "nice" CS properties[1]. Precisely, an MMV system of size $p \times N$ and joint sparsity of order $K$ needs to be solved correctly with that $\mathbf{A}$. In practice, to solve the MMV (3.19), we can make use of existing algorithms from the CS literature, cf. [44–48]. The Introduction and relevant chapters of this book describe various conditions on CS matrices to ensure stable recovery. The dimension requirements of the specific MMV solver in use will impact the number of branches $p$, and consequently the total sampling rate.

---

[1] We comment that most known constructions of "nice" CS matrices involve randomness. In practice, the X-ADC hardware is fixed and defines a deterministic sensing matrix $\mathbf{A}$ for the corresponding IMV system.

The sparse-SI framework can be used, in principle, to reduce the rate of any signal of the form (3.11). In the next section, we treat multiband signals and derive a sub-Nyquist acquisition strategy for this model from the general sparse-SI architecture of Fig. 3.4.

## 3.5    From Theory to Hardware of Multiband Sampling

The prime goal of Xampling is to enable theoretical ideas develop from the math to hardware, to real-world applications. In this section, we study sub-Nyquist sampling of multiband signals in the eyes of a practitioner, aiming to design lowrate sampling hardware. We define the multiband model and propose a union formulation that fits the sparse-SI framework introduced in the previous section. A periodic nonuniform sampling (PNS) solution [19] is then derived from Fig. 3.4. Moving on to practical aspects, we examine frontend bandwidth specifications of commercial ADC devices, and conclude that devices with Nyquist-rate bandwidth are required whenever the ADC is directly connected to a wideband input. Consequently, although PNS as well as the general architecture of Fig. 3.4, enable in principle sub-Nyquist sampling, in practice, high analog bandwidth is necessary, which can be limiting in high-rate applications. To overcome this possible limitation, an alternative scheme, the MWC [20], is presented and analyzed. We conclude our study with a glimpse at circuit aspects that are unique to Xampling systems, as were reported in the circuit design of an MWC prototype hardware [22].

### 3.5.1    Signal Model and Sparse-SI Formulation

The class of multiband signals models a scenario in which $x(t)$ consists of several concurrent RF transmissions. A receiver that intercepts a multiband $x(t)$ sees the typical spectral support that is depicted in Fig. 3.1. We assume that the multiband spectrum contains at most $N$ (symmetric) frequency bands with carriers $f_i$, each of maximal width $B$. The carriers are limited to a maximal frequency $f_{\max}$. The information bands represent analog messages or digital bits transmitted over a shared channel.

When the carrier frequencies $f_i$ are fixed, the resulting signal model can be described as a subspace, and standard demodulation techniques may be used to sample each of the bands at a low rate. A more challenging scenario is when the carriers $f_i$ are unknown. This situation arises, for example, in spectrum sensing for mobile cognitive radio (CR) receivers [23, 51], which aim at utilizing unused frequency regions on an opportunistic basis. Commercialization of CR technology necessitates a spectrum sensing mechanism that can sense a wideband spectrum which consists of several narrowband transmissions, and determines in real time which frequency bands are active.

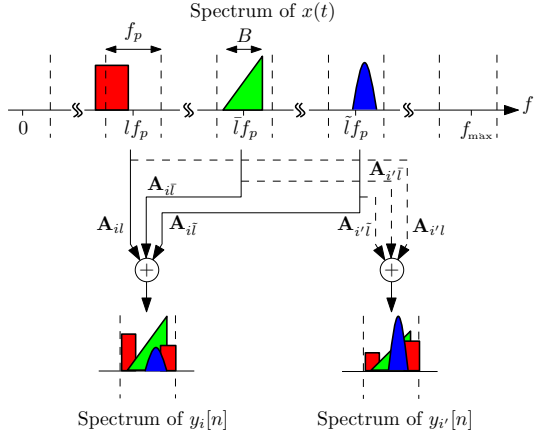

**Figure 3.6:** Spectrum slices of $x(t)$ are overlayed in the spectrum of the output sequences $y_i[n]$. In the example, channels $i$ and $i'$ realize different linear combinations of the spectrum slices centered around $lf_p, \bar{l}f_p, \tilde{l}f_p$. For simplicity, the aliasing of the negative frequencies is not drawn (taken from [22]).

Since each combination of carrier frequencies determines a single subspace, a multiband signal can be described in terms of a union of subspaces. In principle, $f_i$ lies in the continuum $f_i \in [0, f_{\max}]$, so that the union contains infinitely many subspaces. To utilize the sparse-SI framework with finitely many SI generators, a different viewpoint can be used, which treats the multiband model as a finite union of bandpass subspaces, termed spectrum slices [20]. To obtain the finite union viewpoint, the Nyquist range $[-f_{\max}, f_{\max}]$ is conceptually divided into $M = 2L + 1$ consecutive, non-overlapping, slices of individual widths $f_p = 1/T$, such that $M/T \geq f_{\text{NYQ}}$, as depicted in Fig. 3.6. Each spectrum slice represents an SI subspace $\mathcal{V}_i$ of a single bandpass slice. By choosing $f_p \geq B$, we ensure that no more than $2N$ spectrum slices are active, namely contain signal energy. Thus, (3.14) holds with $\mathcal{A}_\lambda$ being the sum over $2N$ SI bandpass subspaces $\mathcal{V}_i$. Consequently, instead of enumerating over the unknown carriers $f_i$, the union is defined over the active bandpass subspaces [16, 19, 20], which can be written in the form (3.11). Note that the conceptual division to spectrum slices does not restrict the band positions; a single band can split between adjacent slices.

Formulating the multiband model with unknown carriers as a sparse-SI problem, we can now apply the sub-Nyquist sampling scheme of Fig. 3.4 to develop an analog CS system for this setting.

### 3.5.2   Analog Compressed Sensing via Nonuniform Sampling

One way to realize the sampling scheme of Fig. 3.4 is through PNS [19]. This strategy is derived from Fig. 3.4 when choosing

$$w_i(t) = \delta(t - c_i T_{\text{NYQ}}), \quad 1 \leq i \leq p, \tag{3.21}$$

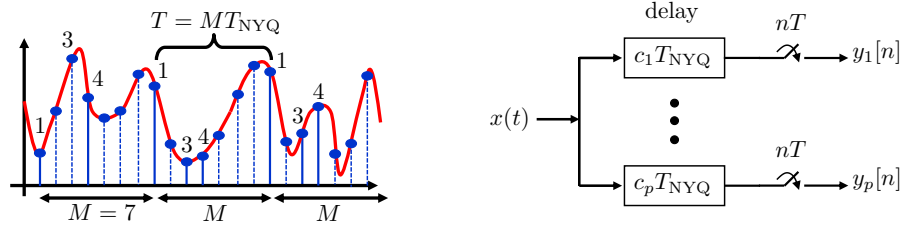

**Figure 3.7:** Periodic nonuniform sampling for sub-Nyquist sensing. In the example, out of $M = 7$ points, only $p = 3$ are active, with time shifts $c_i = 1, 3, 4$.

where $T_{\text{NYQ}} = 1/f_{\text{NYQ}}$ is the Nyquist period, and using a sampling period of $T = MT_{\text{NYQ}}$. Here $c_i$ are integers which select part of the uniform Nyquist grid, resulting in $p$ uniform sequences

$$y_i[n] = x((nM + c_i)T_{\text{NYQ}}). \tag{3.22}$$

The sampling sequences are illustrated in Fig. 3.7. It can be shown that the PNS sequences $y_i[n]$ satisfies an IMV system of the form (3.12) with $d_\ell[n]$ representing the contents of the $\ell$th bandpass slice. The sensing matrix $\mathbf{A}$ in this setting has $i\ell$th entry

$$\mathbf{A}_{i\ell} = e^{j\frac{2\pi}{M}c_i \ell}, \tag{3.23}$$

that is a partial discrete Fourier transform (DFT), obtained by taking only the row indices $c_i$ from the full $M \times M$ DFT matrix. CS properties of partial-DFT matrices are studied in [4], for example.

To recover $x(t)$, we can apply the CTF framework and obtain spectrum blind reconstruction (SBR) of $x(t)$ [19]. Specifically, a frame $\mathbf{Q}$ is computed with (3.20) and is optionally decomposed to another frame $\mathbf{V}$ (to combat noise). Solving (3.19) then indicates the active sequences $d_\ell[n]$, and equivalently estimates the frequency support of $x(t)$ at a coarse resolution of slice width $f_p$. Continuous reconstruction is then obtained by standard lowpass interpolation of the active sequences $d_\ell[n]$ and modulation to the corresponding positions on the spectrum. This procedure is termed SBR4 in [19], where 4 designates that under the choice of $p \geq 4N$ sampling sequences (and additional conditions), this algorithm guarantees perfect reconstruction of a multiband $x(t)$. With the earlier choice $f_p = 1/T \geq B$, the average sampling rate can be as low as $4NB$.

The rate can be further reduced by a factor of 2 exploiting the way a multiband spectra is arranged in spectrum slices. Using several CTF instances, an algorithm reducing the required rate was developed in [19] under the name SBR2, leading to $p \geq 2N$ sampling branches, so that the sampling rate can approach $2NB$. This is essentially the provable optimal rate [19], since regardless of the sampling strategy, theoretic arguments show that $2NB$ is the lowest possible sampling rate for multiband signals with unknown spectrum support [19]. Figure 3.8 depicts recovery performance in Monte Carlo simulations of a (complex-valued) multiband model with $N = 3$ bands, widths $B = 1$ GHz and $f_{\text{NYQ}} = 20$ GHz. Recovery of noisy signals is also simulated in [19]. We demonstrate robustness to noise later on in this section in the context of MWC sampling. The robustness follows from that of the MMV system used for SBR.

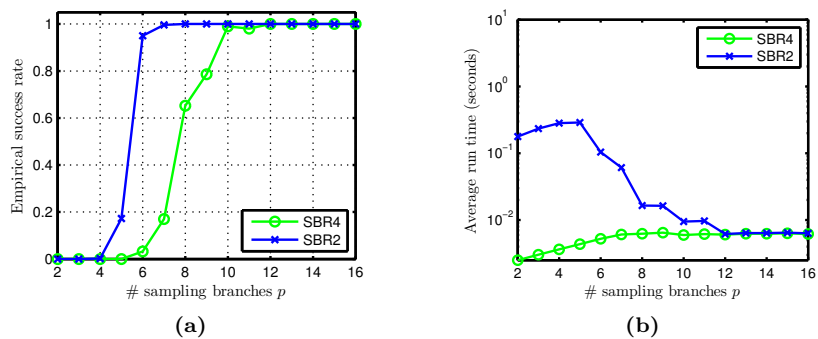

**Figure 3.8:** Comparing algorithms SBR4 and SBR2. (a) Empirical recovery rate for different sampling rates and (b) digital complexity as measured by average run time (taken from [19]).

We note that PNS was utilized for multiband sampling already in classic studies, though the traditional goal was to approach a rate of $NB$ samples/sec. This rate is optimal according to the Landau theorem [52], though achieving it for all input signals is possible only when the spectral support is known and fixed. When the carrier frequencies are unknown, the optimal rate is $2NB$ [19]. Indeed, [11, 53] utilized knowledge of the band positions to design a PNS grid and the required interpolation filters for reconstruction. The approaches in [12, 13] were semi-blind: a sampler design independent of band positions combined with the reconstruction algorithm of [11] which requires exact support knowledge. Other techniques targeted the rate $NB$ by imposing alternative constraints on the input spectrum [21]. Here we demonstrate how analog CS tools [16, 44] can lead to a fully-blind sampling system of multiband inputs with unknown spectra at the appropriate optimal rate [19]. A more thorough discussion in [19] studies the differences between the analog CS method presented here based on [16,19,44] and earlier approaches.

### 3.5.3 Modeling Practical ADC Devices

Analog CS via PNS results in a simple acquisition strategy, which consists of $p$ delay components and $p$ uniform ADC devices. Furthermore, if high sampling rate is not an obstacle and only low processing rates are of interest, then PNS can be simluated by first sampling $x(t)$ at its Nyquist rate and then reducing the rate digitally by discarding some of the samples. Nonuniform topologies of this class are also popular in the design of Nyquist-rate time-interleaved ADC devices, in which case $p = M$ [54, 55].

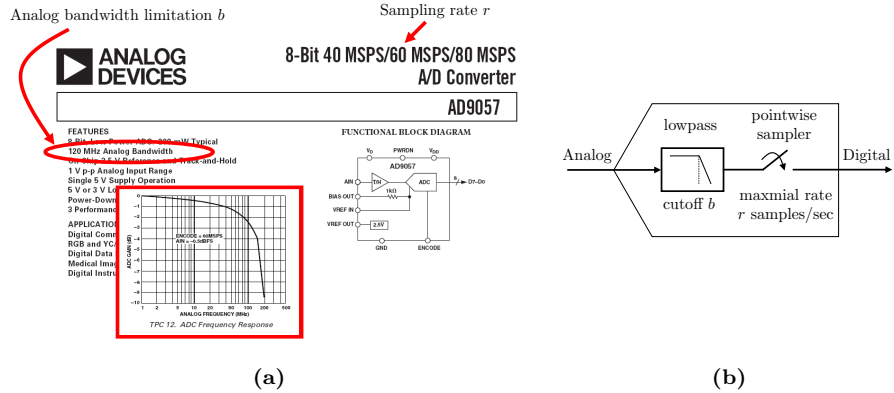

**(a)**             **(b)**

**Figure 3.9:** (a) Datasheet of AD9057 (source: `http://www.analog.com/static/imported-files/data_sheets/AD9057.pdf`). (b) Modeling the inherent bandwidth limitation of the ADC front-end as a lowpass filter preceding pointwise acquisition (taken from [20]).

Realization of a PNS grid with standard ADCs remains simple as long as the input bandwidth is not too high. For high bandwidth signals, PNS is potentially limited, as we now explain by zooming into the drawing of the ADC device of Fig. 3.2. In the signal processing community, an ADC is often modeled as an ideal pointwise sampler that takes snapshots of $x(t)$ at a constant rate of $r$ samples/second. The sampling rate $r$ is the main parameter that is highlighted in the datasheets of popular ADC devices; see online catalogues [56, 57] for many examples.

For most analysis purposes, the first-order model of pointwise acquisition approximates the true ADC operation sufficiently well. Another property of practical devices, also listed in datasheets, is about to play a major role in the UoS settings – the analog bandwidth power $b$. The parameter $b$ measures the $-3$ dB point in the frequency response of the ADC device, which stems from the responses of all circuitries comprising the internal front-end. See the datasheet quote of AD9057 in Fig. 3.9. Consequently, inputs with frequencies up to $b$ Hz can be reliably converted. Any information beyond $b$ is attenuated and distorted. Figure 3.9 depicts an ADC model in which the pointwise sampler is preceded by a lowpass filter with cutoff $b$, in order to take into account the bandwidth limitation [20]. In Xampling, the input signal $x(t)$ belongs to a union set $\mathcal{U}$ which typically has high bandwidth, *e.g.,* multiband signals whose spectrum reaches up to $f_{\max}$ or FRI signals with wideband pulse $h(t)$. This explains the necessity of an analog compression operator $P$ to reduce the bandwidth prior to the actual ADC. The next stage can then employ commercial devices with low analog bandwidth $b$.

The Achilles heel of nonuniform sampling is the pointwise acquisition of a wideband input. While the rate of each sequence $y_i[n]$ is low, namely $f_{\mathrm{NYQ}}/M$, the ADC device still needs to capture a snapshot of a wideband input with frequencies possibly reaching up till $f_{\max}$. In practice, this requires an ADC with front-end bandwidth that reaches the Nyquist rate, which can be challenging in wideband scenarios.

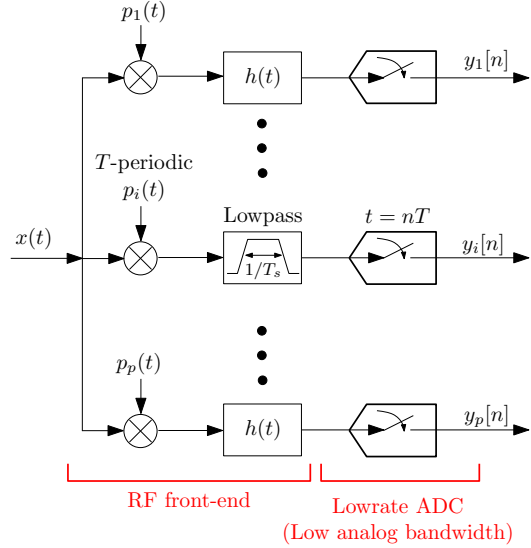

**Figure 3.10:** Block diagram of the modulated wideband converter. The input passes through $p$ parallel branches, where it is mixed with a set of periodic functions $p_i(t)$, lowpass filtered and sampled at a low rate (taken from [22]).

### 3.5.4  Modulated Wideband Converter

To circumvent analog bandwidth issues, an alternative to PNS sensing referred to as the modulated wideband converter (MWC) was developed in [20]. The MWC combines the spectrum slices $d_\ell[n]$ according to the scheme depicted in Fig. 3.10. This architecture allows to implement an effective demodulator without the carrier frequencies being known to the receiver. A nice feature of the MWC is a modular design so that for known carrier frequencies the same receiver can be used with fewer channels or lower sampling rate. Furthermore, by increasing the number of channels or the rate on each channel the same realization can be used for sampling full band signals at the Nyquist rate.

The MWC consists of an analog front-end with $p$ channels. In the $i$th channel, the input signal $x(t)$ is multiplied by a periodic waveform $p_i(t)$ with period $T$, lowpass filtered by an analog filter with impulse response $h(t)$ and cutoff $1/2T$, and then sampled at rate $f_s = 1/T$. The mixing operation scrambles the spectrum of $x(t)$, such that a portion of the energy of all bands appears in baseband. Specifically, since $p_i(t)$ is periodic, it has a Fourier expansion

$$p_i(t) = \sum_{\ell=-\infty}^{\infty} c_{i\ell} e^{j\frac{2\pi}{T}\ell t}. \tag{3.24}$$

In the frequency domain, mixing by $p_i(t)$ is tantamount to convolution between $X(f)$ and the Fourier transform of $p_i(t)$. The latter is a weighted Dirac-comb, with Dirac locations on $f = l/T$ and weights $c_{i\ell}$. Thus, as before, the spectrum is conceptually divided into slices of width $1/T$, represented by the unknown sequences $d_\ell[n]$, and a weighted-sum of these slices is shifted to the origin [20]. The lowpass filter $h(t)$ transfers only the narrowband frequencies up to $f_s/2$ from that mixture to the output sequence $y_i[n]$. The output has the same aliasing pattern that was illustrated in Fig. 3.6. Sensing with the MWC results in the IMV system (3.12) with a sensing matrix $\mathbf{A}$ whose entries are the Fourier expansion coefficients $c_{i\ell}$.

The basic MWC parameter setting is [20]

$$p \geq 4N, \quad f_s = \frac{1}{T} \geq B. \tag{3.25}$$

Using the SBR2 algorithm of [19], the required number of branches is $p \geq 2N$ so that the sampling rate is reduced by a factor of 2 and can approach the minimal rate of $2NB$. Advanced configurations enable additional hardware savings by collapsing the number of branches $p$ by a factor of $q$ at the expense of increasing the sampling rate of each channel by the same factor, ultimately enabling a single-channel sampling system [20]. This property is unique to MWC sensing, since it decouples the aliasing from the actual acquisition.

The periodic functions $p_i(t)$ define a sensing matrix $\mathbf{A}$ with entries $c_{i\ell}$. Thus, as before, $p_i(t)$ need to be chosen such that the resulting $\mathbf{A}$ has "nice" CS properties. In principle, any periodic function with high-speed transitions within the period $T$ can satisfy this requirement. One possible choice for $p_i(t)$ is a sign-alternating function, with $M = 2L+1$ sign intervals within the period $T$ [20]. Popular binary patterns, *e.g.,* Gold or Kasami sequences, are especially suitable for the MWC [58]. Imperfect sign alternations are allowed as long as periodicity is maintained [22]. This property is crucial since precise sign alternations at high speeds are extremely difficult to maintain, whereas simple hardware wirings ensure that $p_i(t) = p_i(t+T)$ for every $t \in \mathbb{R}$ [22]. Another important practical design aspect is that a filter $h(t)$ with nonflat frequency response can be used since a nonideal response can compensated for in the digital domain, using an algorithm developed in [59].

In practical scenarios, $x(t)$ is contaminated by wideband analog noise $e_\text{analog}(t)$ and measurement noise $e_{\ell,\text{meas.}}[n]$ that is added to the compressive sequences $y_\ell[n]$. This results in a noisy IMV system

$$\mathbf{y}[n] = \mathbf{A}(\mathbf{d}[n] + \mathbf{e}_\text{analog}[n]) + \mathbf{e}_\text{meas.}[n] = \mathbf{A}\mathbf{d}[n] + \mathbf{e}_\text{eff.}[n], \tag{3.26}$$

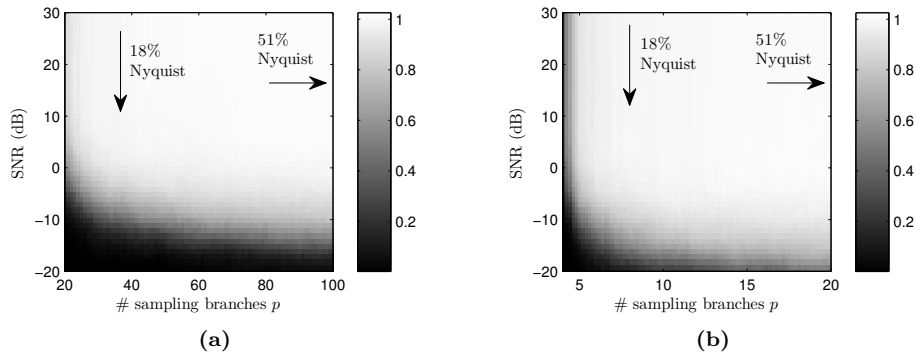

**Figure 3.11:** Image intensity represents percentage of correct recovery of the active slices set $I$, for different number of sampling branches $p$ and under several SNR levels. The collapsing factors are (a) $q = 1$ and (b) $q = 5$. The markers indicate reference points with same total sampling rate $pf_s$ as a fraction of $f_{\text{NYQ}} = 10$ GHz (taken from [20]).

with an effective error term $\mathbf{e}_{\text{eff}}[n]$. This means that noise has the same effects in analog CS as it has in the standard CS framework with an increase in variance due to the term $\mathbf{A}\mathbf{e}_{\text{analog}}[n]$. Therefore, existing algorithms can be used to try and combat the noise. Furthermore, we can translate known results and error guarantees developed in the context of CS to handle noisy analog environments. In particular, as is known in standard CS, the total noise, *i.e.,* in both zero and nonzero locations, is what dictates the behavior of various algorithms and recovery guarantees. Similarly, analog CS systems, such as sparse-SI [16], PNS [19] or MWC [20], aggregate wideband noise power from the entire Nyquist range $[-f_{\max}, f_{\max}]$ into their samples. This is different from standard demodulation that aggregates only in-band noise, since only a specific range of frequencies is shifted to baseband. Nonetheless, as demonstrated below, analog CS methods exhibit robust recovery performance which degrades gracefully as noise levels increase.

Numerical simulations were used in [58] to evaluate the MWC performance in noisy environments. A multiband model with $N = 6$, $B = 50$ MHz and $f_{\text{NYQ}} = 10$ GHz was used to generate inputs $x(t)$, which were contaminated by additive wideband Gaussian noise. An MWC systems with $f_p = 51$ MHz and a varying number $p$ of branches was considered, with sign alternating waveforms of length $M = 195$. Performance of support recovery using CTF is depicted in Fig. 3.11 for various (wideband) signal-to-noise ratio (SNR) levels. Two MWC configurations were tested: a basic version with sampling rate $f_p$ per branch, and an advanced setup with a collapsing factor $q = 5$, in which case each branch samples at rate $qf_p$. The results affirm saving in hardware branches by a factor of 5 while maintaining comparable recovery performance. Signal reconstruction is demonstrated in the next subsection using samples obtained by a hardware MWC prototype.

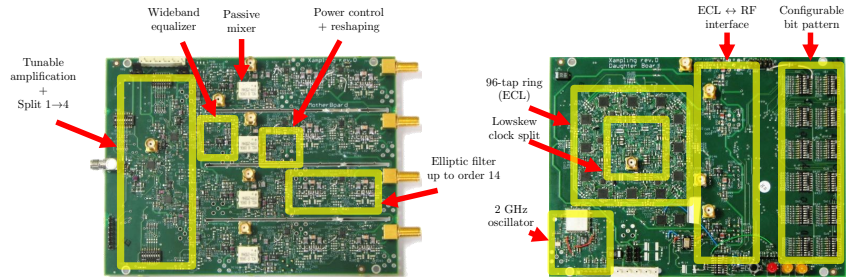

**Figure 3.12:** Hardware realization of the MWC consisting of two circuit boards. The left pane implements $m = 4$ sampling channels, whereas the right pane provides four sign-alternating periodic waveforms of length $M = 108$, derived from different taps of a single shift-register (taken from [22, 60]).

Note that the MWC achieves a similar effect of aliasing bandpass slices to the origin as does the PNS system. However, in contrast to PNS, the MWC accomplishes this goal with analog pre-processing prior to sampling, as proposed in Xampling, which allows the use of standard low-rate ADCs. In other words, the practical aspects of front-end bandwidth motivate a solution which departs from the generic scheme of Fig. 3.4. This is analogous to the advantage of standard demodulation over plain undersampling; both demodulation and undersampling can shift a single bandpass subspace to the origin. However, while undersampling requires an ADC with Nyquist-rate front-end bandwidth, demodulation uses RF technology to interact with the wideband input, thereby requiring only lowrate and low bandwidth ADC devices.

### 3.5.5　Hardware Design

The MWC has been implemented as a board-level hardware prototype [22]. The hardware specifications cover inputs with 2 GHz Nyquist rate and $NB = 120$ MHz spectrum occupation. The prototype has $p = 4$ sampling branches, with total sampling rate of 280 MHz, far below the 2 GHz Nyquist rate. In order to save analog components, the hardware realization incorporates the advanced configuration of the MWC [20] with a collapsing factor $q = 3$. In addition, a single shift-register provides a basic periodic pattern, from which $p$ periodic waveforms are derived using delays, that is, by tapping $p$ different locations of the register. Photos of the hardware are presented in Fig. 3.12.

Several nonordinary RF blocks in the MWC prototype are highlighted in Fig. 3.12. These nonordinary circuitries stem from the unique application of sub-Nyquist sampling as described in detail in [22]. For instance, ordinary analog mixers are specified and commonly used with a pure sinusoid in their oscillator port. The MWC, however, requires simultaneous mixing with the many sinusoids comprising $p_i(t)$. This results in attenuation of the output and substantial non-linear distortion not accounted for in datasheet specifications. To address this challenge, power control, special equalizers and local adjustments on datasheet specifications were used in [22] in order to design the analog acquisition, taking into account the nonordinary mixer behavior due to the periodic mixing.

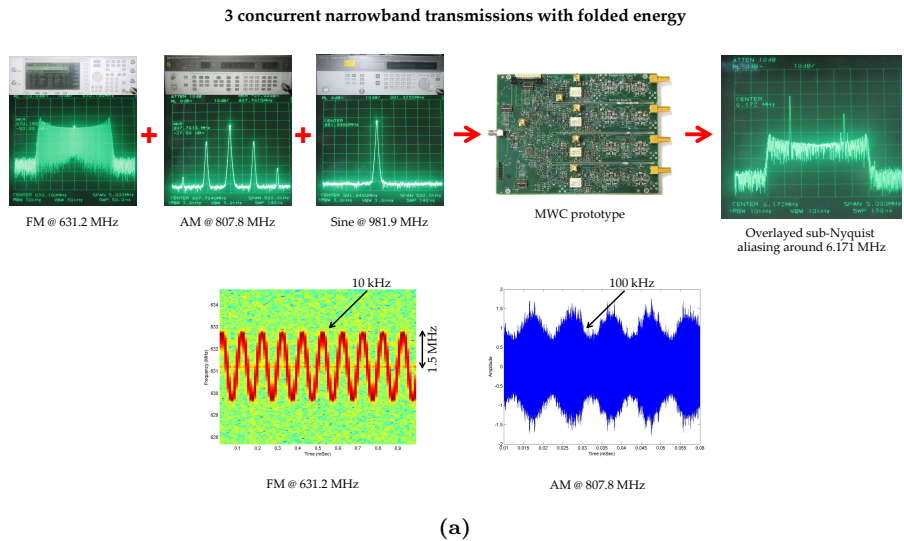

**Figure 3.13:** Three signal generators are combined to the system input terminal. The spectrum of the lowrate samples (first channel) reveals overlapped aliasing at baseband. The recovery algorithm finds the correct carriers and reconstructs the original individual signals (taken from [22]).

Another circuit challenge pertains to generating $p_i(t)$ with 2 GHz alternation rates. The waveforms can be generated either by analog or digital means. Analog waveforms, such as sinusoid, square or sawtooth waveforms, are smooth within the period, and therefore do not have enough transients at high frequencies which is necessary to ensure sufficient aliasing. On the other hand, digital waveforms can be programmed to any desired number of alternations within the period, but require meeting timing constraints on the order of the clock period. For 2 GHz transients, the clock interval $1/f_{\text{NYQ}} = 480$ picosecs leads to tight timing constraints that are difficult to satisfy with existing digital devices. The timing constraints involved in this logic are overcome in [22] by operating commercial devices beyond their datasheet specifications. The reader is referred to [22] for further technical details.

Correct support detection and signal reconstruction in the presence of three narrowband transmissions was verified in [22]. Figure 3.13 depicts the setup of three signal generators that were combined at the input terminal of the MWC prototype: an amplitude-modulated (AM) signal at 807.8 MHz with 100 kHz envelope, a frequency-modulation (FM) source at 631.2 MHz with 1.5 MHz frequency deviation and 10 kHz modulation rate, and a pure sine waveform at 981.9 MHz. Signal powers were set to about 35 dB SNR with respect to the wideband noise that folded to baseband. The carrier positions were chosen so that their aliases overlay at baseband, as the photos in Fig. 3.13 demonstrate. The CTF was executed and detected the correct support set $I$. The unknown carrier frequencies were estimated up to 10 kHz accuracy. In addition, the figure demonstrates correct reconstruction of the AM and FM signal contents. Our lab experiments also indicate an average of 10 millisecond duration for the digital computations, including CTF support detection and carrier estimation. The small dimensions of $\mathbf{A}$ ($12 \times 100$ in the prototype configuration) is what makes the MWC practically feasible from a computational perspective.

The results of Fig. 3.13 connect between theory and practice. The same digital algorithms that were used in the numerical simulations of [20] are successfully applied in [22] on real data, acquired by the hardware. This demonstrates that the theoretical principles are sufficiently robust to accommodate circuit non-idealities, which are inevitable in practice. A video recording of these experiments and additional documentation for the MWC hardware are available at http://webee.technion.ac.il/Sites/People/YoninaEldar/Info/hardware.html. A graphical package demonstrating the MWC numerically is available at http://webee.technion.ac.il/Sites/People/YoninaEldar/Info/software/GUI/MWC_GUI.htm.

The MWC board appears to be the first reported hardware example borrowing ideas from CS to realize a sub-Nyquist sampling system for wideband signals, where the sampling and processing rates are directly proportional to the actual bandwidth occupation and not the highest frequency. Alternative approaches which employ discretization of the analog input are discussed in Section 3.8. The realization of these methods recover signals with Nyquist-rates below 1 MHz, falling outside of the class of wideband samplers. Additionally, the signal representations that result from discretization have size proportional to the Nyquist frequency, leading to recovery problems in the digital domain that are much larger than those posed by the MWC.

### 3.5.6   Sub-Nyquist Signal Processing

A nice feature of the MWC recovery stage is that it interfaces seamlessly with standard DSPs by providing (samples of) the narrowband quadrature information signals $I_i(t), Q_i(t)$ which build the $i$th band of interest

$$s_i(t) = I_i(t)\cos(2\pi f_i t) + Q_i(t)\sin(2\pi f_i t). \qquad (3.27)$$

The signals $I_i(t), Q_i(t)$ could have been obtained by classic demodulation had the carriers $f_i$ been known. In the union settings, with unknown carrier frequencies $f_i$, this capability is provided by a digital algorithm, named Back-DSP, that is developed in [14] and illustrated in Fig. 3.14. The Back-DSP algorithm[2] translates the sequences $\mathbf{d}[n]$ to the narrowband signals $I_i(t), Q_i(t)$ that standard DSP packages expect to receive, thereby providing backward compatibility. Only lowrate computations, proportional to the rate of $I_i(t), Q_i(t)$, are used. Back-DSP first detects the band edges, then separates bands occupying the same slice to distinct sequences and stitches together energy that was split between adjacent slices. Finally, the balanced quadricorrelator [61] is applied in order to estimate the carrier frequencies.

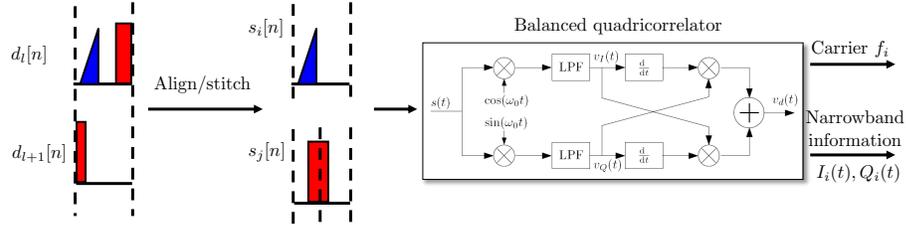

**Figure 3.14:** The flow of information extractions begins with detecting the band edges. The slices are filtered, aligned and stitched appropriately to construct distinct quadrature sequences $s_i[n]$ per information band. The balanced quadricorrelator finds the carrier $f_i$ and extracts the narrowband information signals.

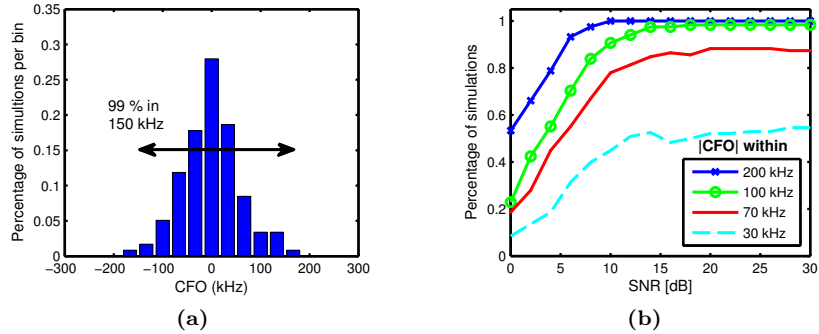

**Figure 3.15:** The distribution of CFO for fixed SNR=10 dB (a). The curves (b) represent the percentage of simulations in which the CFO magnitude is within the specified range (taken from [14]).

Numerical simulations of the Back-DSP algorithm, in a wideband setup similar to the one of Fig. 3.11, evaluated the Back-DSP performance in two aspects. The carrier frequency offset (CFO), estimated vs. true value of $f_i$, is plotted in Fig. 3.15. In most cases, algorithm Back-DSP approaches the true carriers as close as 150 kHz. For reference, the 40 part-per-million (ppm) CFO spec-ifications of IEEE 802.11 standards tolerate 150 kHz offsets for transmissions located around 3.75 GHz [62]. To verify data retrieval, a binary phase-shift keying (BPSK) transmission was generated, such that the band energy splits between two adjacent spectrum slices. A Monte Carlo simulation was used to compute bit error rate (BER) at the output of Back-DSP. Estimated BERs for 3 dB and 5 dB SNR, respectively, are better than $0.77 \cdot 10^{-6}$ and $0.71 \cdot 10^{-6}$. No erroneous bits were detected for SNR of 7 and 9 dB. See [14] for full results.

---

[2] Matlab code is available online at http://webee.technion.ac.il/Sites/People/YoninaEldar/Info/software/FR/FR.htm.

## 3.6  Finite Rate of Innovation Signals

The second class we consider are analog signals in infinite unions of finite-dimensional spaces; these are continuous-time signals that can be characterized by a finite number of coefficients, also termed finite rate of innovaton (FRI) signals as coined by Vetterli et al. [24, 25]. One important problem that is studied in this framework is that of time-delay estimation, in which the input contains several, say $L$, echoes of a known pulse shape $h(t)$, though the echo positions $t_\ell$ and amplitudes $a_\ell$ are unknown [63]. Time-delay estimation is analogous to estimation of frequencies and amplitudes in a mixture of sinusoids. Both problems were widely studied in the classic literature [5, 64–68], with parametric estimation techniques that date back to methods developed by Rife and Boorstyn in 1974 [69] and earlier by David Slepian in the 1950s. The classic approaches focused on improving estimation performance in the digital domain, so that the error in estimating the time-delays, or equivalently the sinusoid frequencies, approaches the optimal defined by the relevant Cramér-Rao bounds. The starting point, however, is discrete samples at the Nyquist rate of the input. The concept of FRI is fundamentally different, as it aims to obtain similar estimates from samples taken at the rate of innovation, namely proportional to $2L$ samples per observation interval, rather than at the typically much-higher rate corresponding to the Nyquist bandwidth of $h(t)$. Chapter 4 in this book provides a comprehensive review of the FRI field. In the present chapter, we focus on Xampling-related aspects, with emphasis on possible hardware configurations for sub-Nyquist FRI acquisition. Recovery algorithms are briefly reviewed for the chapter to be self-contained.

### 3.6.1  Analog Signal Model

As we have seen in Section 3.4, the SI model (3.7) is a convenient way to describe analog signals in infinite-dimensional spaces. We can use a similar approach to describe analog signals that lie within finite-dimensional spaces by restricting the number of unknown gains $a_\ell[n]$ to be finite, leading to the parametrization

$$x(t) = \sum_{\ell=1}^{L} a_\ell h_\ell(t). \tag{3.28}$$

In order to incorporate infiniteness into this model, we assume that each generator $h_\ell(t)$ has an unknown parameter $\alpha_\ell$ associated with it, which can take on values in a continuous interval, resulting in the model

$$x(t) = \sum_{\ell=1}^{L} a_\ell h_\ell(t, \alpha_\ell). \tag{3.29}$$

Each possible choice of the set $\{\alpha_\ell\}$ leads to a different $L$-dimensional subspace of signals $\mathcal{A}_\lambda$, spanned by the functions $\{h(t, \alpha_\ell)\}$. Since $\alpha_\ell$ can take on any value in a given interval, the model (3.29) corresponds to an infinite union of finite dimensional subspaces (i.e., $|\Lambda| = \infty$), where each subspace $\mathcal{A}_\lambda$ in (3.3) contains those analog signals corresponding to a particular configuration of $\{\alpha_\ell\}_{\ell=1}^{L}$.

An important example of (3.29) is when $h_\ell(t, \alpha_\ell) = h(t - t_\ell)$ for some unknown time delay $t_\ell$, leading to a stream of pulses

$$x(t) = \sum_{\ell=1}^{L} a_\ell h(t - t_\ell). \tag{3.30}$$

Here $h(t)$ is a known pulse shape and $\{t_\ell, a_\ell\}_{\ell=1}^{L}$, $t_\ell \in [0, \tau)$, $a_\ell \in \mathbb{C}$, $\ell = 1 \ldots L$ are unknown delays and amplitudes. This model was introduced by Vetterli et al. [24,25] as a special case of signals having a finite number of degrees of freedom per unit time, termed FRI signals. Our goal is to sample $x(t)$ and reconstruct it from a minimal number of samples. Since in FRI applications the primary interest is in pulses which have small time-support, the required Nyquist rate can be very high. Bearing in mind that the pulse shape $h(t)$ is known, there are only $2L$ degrees of freedom in $x(t)$, and therefore, we expect the minimal number of samples to be $2L$, much lower than the number of samples resulting from Nyquist rate sampling.

### 3.6.2    Compressive Signal Acquisition

To date, there are no general acquisition methods for signals of the form (3.29), while there are known solutions to various instances of (3.30). We begin by focusing on a simpler version of the problem, in which the signal $x(t)$ of (3.30) is repeated periodically leading to the model

$$x(t) = \sum_{m \in \mathbb{Z}} \sum_{\ell=1}^{L} a_\ell h(t - t_\ell - m\tau), \tag{3.31}$$

where $\tau$ is a known period. This periodic setup is easier to treat because we can exploit the properties of the Fourier series representation of $x(t)$ due to the periodicity. The dimensionality and number of subspaces included in the model (3.3) remain unchanged.

The key to designing an efficient X-ADC stage for this model is in identifying the connection to a standard problem in signal processing: the retrieval of the frequencies and amplitudes of a sum of sinusoids. The Fourier series coefficients $X[k]$ of the periodic pulse stream $x(t)$ are actually a sum of complex exponentials, with amplitudes $\{a_\ell\}$, and frequencies directly related to the unknown time-delays [24]:

$$X[k] = \frac{1}{\tau} H(2\pi k/\tau) \sum_{\ell=1}^{L} a_\ell e^{-j2\pi k t_\ell/\tau}, \qquad (3.32)$$

where $H(\omega)$ is the CTFT of the pulse $h(t)$. Therefore, once the Fourier coefficients are known, the unknown delays and amplitudes can be found using standard tools developed in the context of array processing and spectral estimation [24, 70]. For further details see Chapter 4 in this book. Our focus here is on how to obtain the Fourier coefficients $X[k]$ efficiently from $x(t)$.

There are several X-ADC operators $P$ which can be used to obtain the Fourier coefficients from time-domain samples of the signal. One choice is to set $P$ to be a lowpass filter, as suggested in [24]. The resulting reconstruction requires $2L+1$ samples and therefore presents a near-critical sampling scheme. A general condition on the sampling kernel $s(t)$ that allows obtaining the Fourier coefficients was derived in [27]: its CTFT $S(\omega)$ should satisfy

$$S(\omega) = \begin{cases} 0, & \omega = 2\pi k/\tau, k \notin \mathcal{K} \\ \text{nonzero}, & \omega = 2\pi k/\tau, k \in \mathcal{K} \\ \text{arbitrary}, & \text{otherwise}, \end{cases} \qquad (3.33)$$

where $\mathcal{K}$ is a set of $2L$ consecutive indices such that $H\left(\frac{2\pi k}{\tau}\right) \neq 0$ for all $k \in \mathcal{K}$. The resulting X-ADC consists of a filter with a suitable impulse response $s(t)$ followed by a uniform sampler.

A special class of filters satisfying (3.33) are Sum of Sincs (SoS) in the frequency domain [27], which lead to compactly supported filters in the time domain. These filters are given in the Fourier domain by

$$G(\omega) = \frac{\tau}{\sqrt{2\pi}} \sum_{k \in \mathcal{K}} b_k \operatorname{sinc}\left(\frac{\omega}{2\pi/\tau} - k\right), \qquad (3.34)$$

where $b_k \neq 0$, $k \in \mathcal{K}$. It is easy to see that this class of filters satisfies (3.33) by construction. Switching to the time domain leads to

$$g(t) = \operatorname{rect}\left(\frac{t}{\tau}\right) \sum_{k \in \mathcal{K}} b_k e^{j2\pi k t/\tau}. \qquad (3.35)$$

For the special case in which $\mathcal{K} = \{-p, \ldots, p\}$ and $b_k = 1$,

$$g(t) = \operatorname{rect}\left(\frac{t}{\tau}\right) \sum_{k=-p}^{p} e^{j2\pi k t/\tau} = \operatorname{rect}\left(\frac{t}{\tau}\right) D_p(2\pi t/\tau), \qquad (3.36)$$

where $D_p(t)$ denotes the Dirichlet kernel.

While periodic streams are mathematically convenient, finite pulse streams of the form (3.30) are ubiquitous in real world applications. A finite pulse stream can be viewed as a restriction of a periodic FRI signal to a single period. As long as the analog preprocessing $P$ does not involve values of $x(t)$ outside the observation interval $[0, \tau]$, this implies that sampling and reconstruction methods developed for the periodic case also apply to finite settings. Treating time-limited signals with lowpass $P$, however, may be difficult since it has infinite time support, beyond the interval $[0, \tau]$ containing the finite pulse stream. Instead, we can choose fast-decaying sampling kernels or SoS filters such as (3.35) that have compact time support $\tau$ by construction.

To treat the finite case, a Gaussian sampling kernel was proposed in [24]; however, this method is numerically unstable since the samples are multiplied by a rapidly diverging or decaying exponent. As an alternative, we may use compactly supported sampling kernels for certain classes of pulse shapes based on splines [25]; this enables obtaining moments of the signal rather than its Fourier coefficients. These kernels have several advantages in practice as detailed in the next chapter. The moments are then processed in a similar fashion (see the next subsection for details). However, this approach is unstable for high values of $L$ [25]. To improve robustness, the SoS class is extended to the finite case by exploiting the compact support of the filters [27]. This approach exhibits superior noise robustness when compared to the Gaussian and spline methods, and can be used for stable reconstruction even for very high values of $L$, e.g., $L = 100$.

The model of (3.30) can be further extended to the infinite stream case, in which

$$x(t) = \sum_{\ell \in \mathbb{Z}} a_\ell h(t - t_\ell), \quad t_\ell, a_\ell \in \mathbb{R}. \qquad (3.37)$$

Both [25] and [27] exploit the compact support of their sampling filters, and show that under certain conditions the infinite stream may be divided into a series of finite problems, which are solved independently with the existing finite algorithm. However, both approaches operate far from the rate of innovation, since proper spacing is required between the finite streams in order to allow the reduction stage, mentioned earlier. In the next section we consider a special case of (3.37) in which the time delays repeat periodically (but not the amplitudes). As we will show in this special case, efficient sampling and recovery is possible even using a single filter, and without requiring the pulse $h(t)$ to be time limited.

An alternative choice of analog compression operator $P$ to enable recovery of infinite streams of pulses is to introduce multichannel sampling schemes. This approach was first considered for Dirac streams, where a successive chain of integrators allows obtaining moments of the signal [71]. Unfortunately, the method is highly sensitive to noise. A simple sampling and reconstruction scheme consisting of two channels, each with an RC circuit, was presented in [72] for the special case where there is no more than one Dirac per sampling period. A more general multichannel architecture that can treat a broader class of pulses, while being much more stable, is depicted in Fig. 3.16 [28]. The system is very similar to the MWC presented in the previous section, and as such it also complies with the general Xampling architecture. In each channel of this X-ADC, the signal is mixed with a modulating waveform $s_\ell(t)$, followed by an integrator, resulting in a mixture of the Fourier coefficients of the signal. By correct choice of the mixing coefficients, the Fourier coefficients may be extracted from the samples by a simple matrix inversion. This method exhibits superior noise robustness over the integrator chain method [71] and allows for more general compactly supported pulse-shapes. A recent method studied multi-channel sampling for analog signals comprised of several, possibly overlapping, finite duration pulses with unknown shapes and time positions [73].

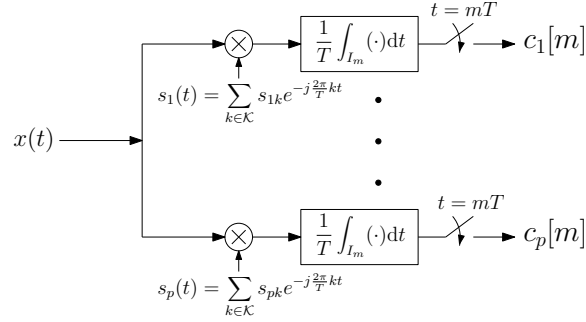

**Figure 3.16:** Extended sampling scheme using modulating waveforms for an infinite pulse stream (taken from [28]).

From a practical hardware perspective it is often more convenient to implement the multichannel scheme rather than a single-channel acquisition with an analog filter that satisfies the SoS structure. It is straightforward to show that the SoS filtering approach can also be implemented in the form of Fig. 3.16 with coefficient matrix $\mathbf{S} = \mathbf{Q}$ where $\mathbf{Q}$ is chosen according to the definition following (3.38), for the SoS case. We point out that the multichannel architecture of Fig. 3.16 can be readily implemented using the MWC prototype hardware. Mixing functions $s_\ell(t)$ comprised of finitely many sinusoids can be obtained by properly filtering a general periodic waveform. Integration over $T$ is a first order lowpass filter which can be assembled in place of the typically higher-order filter of the MWC system [60].

### 3.6.3  Recovery Algorithms

In both the single-channel and multichannel approaches, recovery of the unknown delays and amplitudes proceeds according to Xampling by detecting the parameters $t_\ell$ that identify the signal subspace. The approach consists of two steps. First, the vector of samples $\mathbf{c}$ is related to the Fourier coefficients vector $\mathbf{x}$ through a $p \times |\mathcal{K}|$ mixing matrix $\mathbf{Q}$, as

$$\mathbf{c} = \mathbf{Q}\mathbf{x}. \tag{3.38}$$

Here $p \geq 2L$ represents the number of samples. When using the SoS approach with a filter $S(\omega)$, $\mathbf{Q} = \mathbf{VS}$ where $\mathbf{S}$ is a $p \times p$ diagonal matrix with diagonal elements $S^*(-2\pi\ell/\tau)$, $1 \leq \ell \leq p$, and $\mathbf{V}$ is a $p \times |\mathcal{K}|$ Vandermonde matrix with $\ell$th element given by $e^{j2\pi\ell T/\tau}$, $1 \leq \ell \leq p$, where $T$ denotes the sampling period. For the multichannel architecture of Fig. 3.16, $\mathbf{Q}$ consists of the modulation coefficients $s_{\ell k}$. The Fourier coefficients $\mathbf{x}$ can be obtained from the samples as

$$\mathbf{x} = \mathbf{Q}^\dagger \mathbf{c}. \tag{3.39}$$

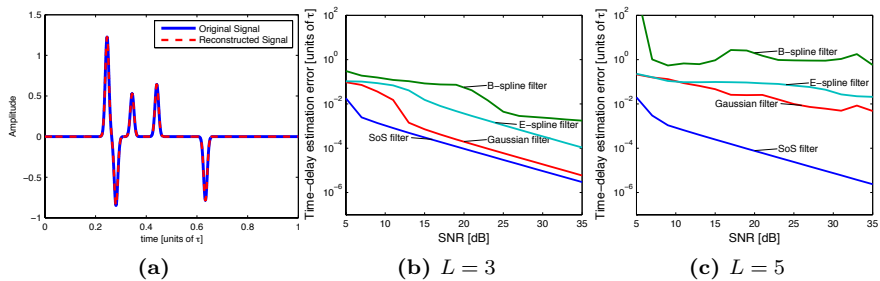

**Figure 3.17:** Performance comparison of finite pulse stream recovery using Gaussian, B-spline, E-spline, and SoS sampling kernels. (a) Reconstructed signal using SoS filters vs. original one. The reconstruction is exact to numerical precision. (b) $L = 3$ Dirac pulses are present, (c) $L = 5$ pulses (taken from [27]).

The unknown parameters $\{t_\ell, a_\ell\}_{\ell=1}^L$ are then recovered from $\mathbf{x}$ using standard spectral estimation tools, e.g. the annihilating filter method (see [24, 70] and the next chapter for details). These techniques can operate with as low as $2L$ Fourier coefficients. When a larger number of samples are available, alternative techniques that are more robust to noise can be used, such as the matrix-pencil method [74], and the Tufts and Kumaresan technique [75]. In Xampling terminology, these methods detect the input subspace, analogous to the role that CS plays in the CTF block for sparse-SI or multiband unions.

Reconstruction results for the sampling scheme using an SoS filter (3.34) with $b_k = 1$ are depicted in Fig. 3.17. The original signal consists of $L = 5$ Gaussian pulses, and $N = 11$ samples were used for reconstruction. The reconstruction is exact to numerical precision. A comparison of the performance of various methods in the presence of noise is depicted in Fig. 3.17 for a finite stream consisting of 3 and 5 pulses. The pulse-shape is a Dirac delta, and white gaussian noise is added to the samples with a proper level in order to reach the desired SNR for all methods. All approaches operate using $2L + 1$ samples. The results affirm stable recovery when using SoS filters. Chapter 4 of this book reviews in detail FRI recovery in the presence of noise [76] and outlines potential applications in superresolution imaging [77], ultrasound [27] and radar imaging [29].

## 3.7   Sequences of Innovation Signals

The conventional SI setting (3.7) treats a single input subspace spanned by the shifts of $N$ given generators $h_\ell(t)$. Combining the SI setting (3.7) and the time uncertainties of Section 3.6, we now incorporate structure by assuming that each generator $h_\ell(t)$ is given up to some unknown parameter $\alpha_\ell$ associated with it, leading to an infinite union of infinite-dimensional spaces. As with its finite counterpart, there is currently no general sampling framework available to treat such signals. Instead, we focus on a special time-delay scenario of this model for which efficient sampling techniques have been developed.

### 3.7.1   Analog Signal Model

An interesting special case of the general model (3.29) is when $h_\ell(t) = h(t)$ and $\alpha_\ell = t_\ell$ represent unknown delays, leading to [26, 28, 29]

$$x(t) = \sum_{n \in \mathbb{Z}} \sum_{\ell=1}^{L} a_\ell[n] h(t - t_\ell - nT), \tag{3.40}$$

where $\mathbf{t} = \{t_\ell\}_{\ell=1}^{L}$ is a set of unknown time delays contained in the time interval $[0, T)$, $\{a_\ell[n]\}$ are arbitrary bounded energy sequences, presumably representing lowrate streams of information, and $h(t)$ is a known pulse shape. For a given set of delays $\mathbf{t}$, each signal of the form (3.40) lies in an SI subspace $\mathcal{A}_\lambda$, spanned by $L$ generators $\{h(t - t_\ell)\}_{\ell=1}^{L}$. Since the delays can take on any values in the continuous interval $[0, T)$, the set of all signals of the form (3.40) constitutes an infinite union of SI subspaces, i.e., $|\Lambda| = \infty$. Additionally, since any signal has parameters $\{a_\ell[n]\}_{n \in \mathbb{Z}}$, each of the $\mathcal{A}_\lambda$ subspaces has infinite cardinality. This model generalizes (3.37) with time delays that repeat periodically, where (3.40) allows the pulse shapes to have infinite support.

### 3.7.2   Compressive Signal Acquisition

To obtain a Xampling system, we follow a similar approach to that in Section 3.4, which treats a structured SI setting where there are $N$ possible generators. The difference though is that in this current case there are infinitely many possibilities. Therefore, we replace the CTF detection in the X-DSP of Fig. 3.4 with a detection technique that supports this continuity: we will see that the ESPRIT method essentially replaces the CTF block [6].

A sampling and reconstruction scheme for signals of the form (3.40) is depicted in Fig. 3.18 [26]. The analog compression operator $P$ is comprised of $p$ parallel sampling channels, where $p = 2L$ is possible under mild conditions on the sampling filters [26]. In each channel, the input signal $x(t)$ is filtered by a band-limited sampling kernel $s_\ell^*(-t)$ with frequency support contained in an interval of width $2\pi p/T$, followed by a uniform sampler operating at a rate of $1/T$, thus providing the sampling sequence $c_\ell[n]$. Note that just as in the MWC (Section 3.5.4), the sampling filters can be collapsed to a single filter whose output is sampled at $p$ times the rate of a single channel. In particular, acquisition can be as simple as a single channel with a lowpass filter followed by a uniform sampler. Analog compression of (3.40) is obtained by spreading out the energy of the signal in time, in order to capture all vital information with the narrow range $2\pi p/T$ of frequencies. To understand the importance of this stage, consider the case where $g(t) = \delta(t)$ and there are $L = 2$ Diracs per period of $T = 1$, as illustrated in Fig. 3.19(a). We use a sampling scheme consisting of a complex bandpass filterbank with 4 channels, each with width $2\pi/T$. In Fig. 3.19(b) to (d), the outputs of the first 3 sampling channels are shown. It can be seen that the sampling kernels "smooth" the short pulses (Diracs in this example) in the time domain so that even when the sampling rate is low, the samples contain signal information. In contrast, if the input signal was sampled directly, then most of the samples would be zero.

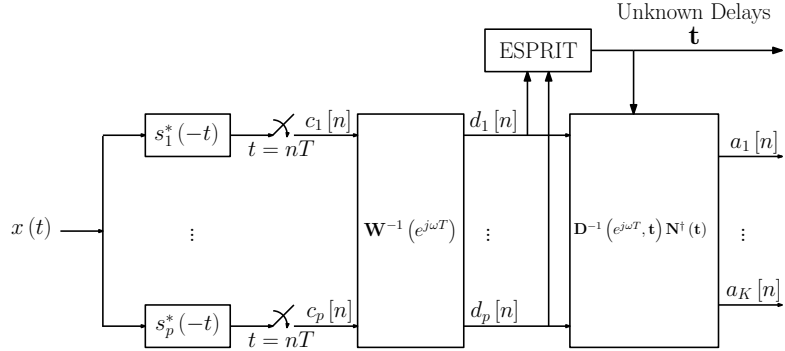

**Figure 3.18:** Sampling and reconstruction scheme for signals of the form (3.40) (taken from [26])

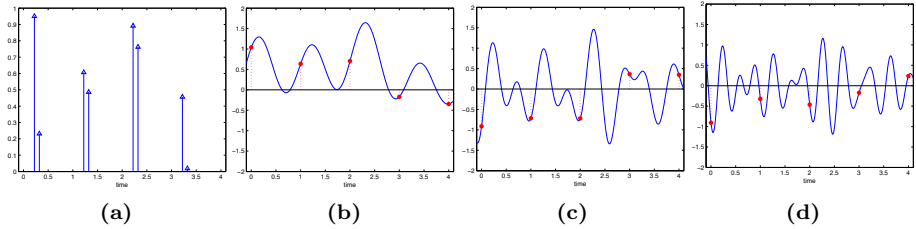

**Figure 3.19:** Stream of Diracs. (a) $L = 2$ Diracs per period $T = 1$. (b)-(d) The outputs of the first three sampling channels, the dashed lines denote the sampling instances (taken from [26]).

### 3.7.3 Recovery Algorithms

To recover the signal from the samples, a properly designed digital filter correction bank, whose frequency response in the DTFT domain is given by $\mathbf{W}^{-1}(e^{j\omega T})$, is applied to the sampling sequences in a manner similar to (3.10). The matrix $\mathbf{W}(e^{j\omega T})$ depends on the choice of the sampling kernels $s_\ell^*(-t)$ and the pulse shape $h(t)$. Its entries are defined for $1 \leq \ell, m \leq p$ as

$$\mathbf{W}\left(e^{j\omega T}\right)_{\ell,m} = \frac{1}{T} S_\ell^*(\omega + 2\pi m/T) H(\omega + 2\pi m/T). \qquad (3.41)$$

After the digital correction stage, it can be shown that the corrected sample vector $\mathbf{d}[n]$ is related to the unknown amplitudes vector $\mathbf{a}[n] = \{a_\ell[n]\}$ by a Vandermonde matrix which depends on the unknown delays [26]. Therefore, subspace detection can be performed by exploiting known tools from the direction of arrival [78] and spectral estimation [70] literature to recover the delays $\mathbf{t} = \{t_1, \ldots, t_L\}$, such as the well-known ESPRIT algorithm [6]. Once the delays are determined, additional filtering operations are applied on the samples to recover the information sequences $a_\ell[n]$. In particular, referring to Fig. 3.18, the matrix $\mathbf{D}$ is a diagonal matrix with diagonal elements equal to $e^{-j\omega t_k}$, and $\mathbf{N}(\mathbf{t})$ is a Vandermonde matrix with elements $e^{-j2\pi m t_k/T}$.

In our setting, the ESPRIT algorithm consists of the following steps:

1. Construct the correlation matrix $\mathbf{R}_{dd} = \sum_{n \in \mathbb{Z}} \mathbf{d}[n] \mathbf{d}^H[n]$.
2. Perform an SVD decomposition of $\mathbf{R}_{dd}$ and construct the matrix $\mathbf{E}_s$ consisting of the $L$ singular vectors associated with the non-zero singular values in its columns.
3. Compute the matrix $\mathbf{\Phi} = \mathbf{E}_{s\downarrow}^\dagger \mathbf{E}_{s\uparrow}$. The notations $\mathbf{E}_{s\downarrow}$ and $\mathbf{E}_{s\uparrow}$ denote the sub matrices extracted from $\mathbf{E}_s$ by deleting its last/first row respectively.
4. Compute the eigenvalues of $\mathbf{\Phi}$, $\lambda_i, i = 1, 2, \ldots, L$.
5. Retrieve the unknown delays by $t_i = -\frac{T}{2\pi} \arg(\lambda_i)$.

In general, the number of sampling channels $p$ required to ensure unique recovery of the delays and sequences using the proposed scheme has to satisfy $p \geq 2L$ [26]. This leads to a minimal sampling rate of $2L/T$. For certain signals, the sampling rate can be reduced even further to $(L+1)/T$ [26]. Interestingly, the minimal sampling rate is not related to the Nyquist rate of the pulse $h(t)$. Therefore, for wideband pulse shapes, the reduction in rate can be quite substantial. As an example, consider the setup in [79], used for characterization of ultra-wide band wireless indoor channels. Under this setup, pulses with bandwidth of $W = 1\mathrm{GHz}$ are transmitted at a rate of $1/T = 2\mathrm{MHz}$. Assuming that there are 10 significant multipath components, we can reduce the sampling rate down to 40MHz compared with the 2GHz Nyquist rate.

We conclude by noting that the approach of [26] imposes only minimal conditions on the possible generator $h(t)$ in (3.40), so that in principle almost arbitrary generators can be treated according to Fig. 3.18, including $h(t)$ with unlimited time support. As mentioned earlier, implementing this sampling strategy can be as simple as collapsing the entire system to a single channel that consists of a lowpass filter and a uniform sampler. Reconstruction, however, involves a $p \times p$ bank of digital filters $\mathbf{W}^{-1}(e^{j\omega T})$, which can be computationally demanding. In scenarios with time-limited $h(t)$ sampling with the multichannel scheme of Fig. 3.16 can be more convenient, since digital filtering is not required so that ESPRIT is applied directly on the samples [28].

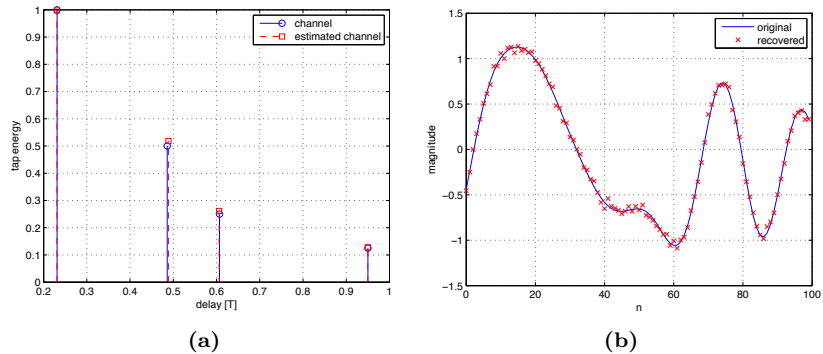

**Figure 3.20:** Channel estimation with $p = 5$ sampling channels, and SNR=20dB. (a) Delay recovery. (b) Recovery of the time-varying gain coefficient of the first path (taken from [26]).

### 3.7.4 Applications

Problems of the form (3.40) appear in a variety of different settings. For example, the model (3.40) can describe multipath medium identification problems, which arise in applications such as radar [80], underwater acoustics [81], wireless communications [82], and more. In this context, pulses with known shape are transmitted through a multipath medium, which consists of several propagation paths, at a constant rate. As a result the received signal is composed of delayed and weighted replicas of the transmitted pulses. The delays $t_\ell$ represent the propagation delays of each path, while the sequences $a_\ell[n]$ describe the time-varying gain coefficient of each multipath component.

An example of multipath channel identification is shown in Fig. 3.20. The channel consists of four propagation paths and is probed by pulses at a rate of $1/T$. The output is sampled at a rate of $5/T$, with white gaussian noise with SNR of 20dB added to the samples. Fig. 3.20 demonstrates recovery of the propagations delays, and the time-varying gain coefficients, from low rate samples corrupted by noise. This is essentially a combination of X-ADC and X-DSP, where the former is used to reduce the sampling rate, while the latter is responsible for translating the compressed sample sequences $c_\ell[n]$ to the set of low rate streams $d_\ell[n]$, which convey the actual information to the receiver. In this example the scheme of Fig. 3.18 was used with a bank of ideal band-pass filters covering consecutive frequency bands:

$$S_\ell(\omega) = \begin{cases} T, & \omega \in \left[(\ell-1)\frac{2\pi}{T}, \ell\frac{2\pi}{T}\right] \\ 0, & \text{otherwise.} \end{cases} \qquad (3.42)$$

As can be seen, even in the presence of noise, the channel is recovered almost perfectly from low rate samples. Applications to radar are explored in Chapter 4 and later on in Section 3.8.5.

## 3.8  Union Modeling vs. Finite Discretization

The approach we have been describing so far treats analog signals by taking advantage of a UoS model, where the inherent infiniteness of the analog signal enters either through the dimensions of the underlying subspaces $\mathcal{A}_\lambda$, the cardinality of the union $|\Lambda|$ or both. An alternative strategy is to assume that the analog signal has some finite representation to begin with, *i.e.*, that both $\Lambda$ and $\mathcal{A}_\lambda$ are finite. Sampling in this case can be readily mapped to a standard underdetermined CS system $\mathbf{y} = \mathbf{A}\mathbf{x}$ (that is with a single vector of unknowns rather than infinitely many as in the IMV setting).

The methods we review in this section treat continuous signals that have an underlying finite parameterization: the RD [30] and quantized CS radar [31–33]. In addition to surveying [30–33], we examine the option of applying sampling strategies developed for finite settings on general analog models with infinite cardinalities. To address this option, we compare hardware and digital complexities of the RD and MWC systems when treating multiband inputs, and imaging performance of quantized [31–33] vs. analog radar [29]. In order to obtain a close approximation to union modeling, a sufficiently dense discretization of the input is required, which in turn can degrade performance in various practical metrics. Thus, whilst methods such as [30–33] are effective for the models for which they were developed, their application to general analog signals, presumably by discretization, may limit the range of signal classes that can be treated.

### 3.8.1  Random Demodulator

The RD approach treats signals consisting of a discrete set of harmonic tones with the system that is depicted in Fig. 3.21 [30].

**Signal model.** A multitone signal $f(t)$ consists of a sparse combination of integral frequencies:

$$f(t) = \sum_{\omega \in \Omega} a_\omega e^{j2\pi\omega t}, \qquad (3.43)$$

where $\Omega$ is a finite set of $K$ out of an even number $Q$ of possible harmonics

$$\Omega \subset \{0, \pm\Delta, \pm 2\Delta, \cdots, \pm(0.5Q-1)\Delta, 0.5Q\Delta\}. \qquad (3.44)$$

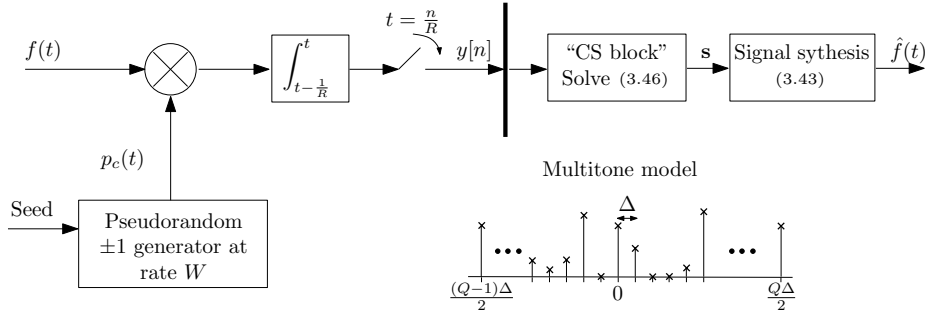

**Figure 3.21:** Block diagram of the random demodulator (taken from [30]).

The model parameters are the tone spacing $\Delta$, number of active tones $K$ and grid length $Q$. The Nyquist rate is $Q\Delta$. Whenever normalized, $\Delta$ is omitted from formulas under the convention that all variables take nominal values (*e.g.,* $R = 10$ instead of $R = 10$ Hz).

**Sampling.** The input signal $f(t)$ is mixed by a pseudorandom chipping sequence $p_c(t)$ which alternates at a rate of $W$. The mixed output is then integrated and dumped at a constant rate $R$, resulting in the sequence $y[n]$, $1 \leq n \leq R$. The development in [30] uses the following parameter setup

$$\Delta = 1, \quad W = Q, \quad R \in \mathbb{Z} \text{ such that } \frac{W}{R} \in \mathbb{Z}. \tag{3.45}$$

It was proven in [30] that if $W/R$ is an integer and (3.45) holds, then the vector of samples $\mathbf{y} = [y[1], \ldots, y[R]]^T$ can be written as

$$\mathbf{y} = \mathbf{\Phi x}, \quad \mathbf{x} = \mathbf{F s}, \quad \|\mathbf{s}\|_0 \leq K. \tag{3.46}$$

The matrix $\mathbf{\Phi}$ has dimensions $R \times W$, effectively capturing the mechanism of integration over $W/R$ Nyquist intervals, where the polarity of the input is flipped on each interval according to the chipping function $p_c(t)$. See Fig. 3.23(a) in the sequel for further details on $\mathbf{\Phi}$. The $W$-squared DFT matrix $\mathbf{F}$ accounts for the sparsity in the frequency domain. The vector $\mathbf{s}$ has $Q$ entries $s_\omega$ which are up to a constant scaling from the corresponding tone amplitudes $a_\omega$. Since the signal has only $K$ active tones, $\|\mathbf{s}\|_0 \leq K$.

**Reconstruction.** Equation (3.46) is an underdetermined system that can be solved with existing CS algorithms, *e.g.,* $\ell_1$-minimization or greedy methods. As before, a "nice" CS matrix $\mathbf{\Phi}$ is required in order to solve (3.46) with sparsity order $K$ efficiently with existing polynomial-time algorithms. In Fig. 3.21, the CS block refers to solving (3.46) with a polynomial-time CS algorithm and a "nice" $\mathbf{\Phi}$, which requires a sampling rate on the order of [30]

$$R \approx 1.7 K \log(W/K + 1). \tag{3.47}$$

Once the sparse $\mathbf{s}$ is found, the amplitudes $a_\omega$ are determined from $s_\omega$ by constant scaling, and the output $\hat{f}(t)$ is synthesized according to (3.43).

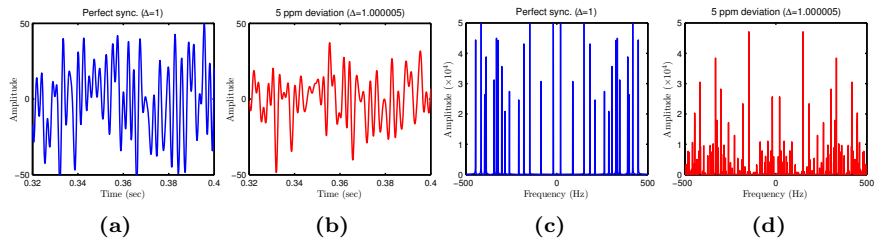

**Figure 3.22:** Effects of non-integral tones on the output of the random demodulator. Panels (a),(b) plot the recovered signal in the time domain. The frequency contents are compared in panels (c),(d) (taken from [14]).

### 3.8.2  Finite-Model Sensitivity

The RD system is sensitive to inputs with tones slightly displaced from the theoretical grid, as was indicated by several studies [14,83,84]. For example, [14] repeated the developments of [30] for an unnormalized multitone model, with $\Delta$ as a free parameter and $W, R$ that are not necessarily integers. The measurements still obey the underdetermined system (3.46) as before, where now [14]

$$W = Q\Delta, \quad R = N_R\Delta, \quad \frac{W}{R} \in \mathbb{Z}, \tag{3.48}$$

and $N_R$ is the number of samples taken by the RD. The equalities in (3.48) imply that the rates $W, R$ need to be perfectly synchronized with the tones spacing $\Delta$. If (3.48) does not hold, either due to hardware imperfections so that the rates $W, R$ deviate from their nominal values, or due to model mismatch so that the actual spacing $\Delta$ is different than what was assumed, then the reconstruction error grows high.

The following toy-example demonstrates this sensitivity. Let $W = 1000, R = 100$ Hz, with $\Delta = 1$ Hz. Construct $f(t)$ by drawing $K = 30$ locations uniformly at random on the tones grid and normally-distributed amplitudes $a_\omega$. Basis pursuit gave exact recovery $\hat{f}(t) = f(t)$ for $\Delta = 1$. For 5 part-per-million (ppm) deviation in $\Delta$ the squared-error reached 37%:

$$\Delta = 1 + 0.000005 \quad \rightarrow \quad \frac{\|f(t) - \hat{f}(t)\|^2}{\|f(t)\|^2} = 37\%. \tag{3.49}$$

Figure 3.22 plots $f(t)$ and $\hat{f}(t)$ in time and frequency, revealing many spurious tones due to the model mismatch. The equality $W = Q$ in the normalized setup (3.45) hints at the required synchronization, though the dependency on the tones spacing is implicit since $\Delta = 1$. With $\Delta \neq 1$, this issue appears explicitly.

The sensitivity that is demonstrated in Fig. 3.22 is a source of error already in the finite multitone setting (3.43). The implication is that utilizing the RD for the counterpart problem of sampling multiband signals with continuous spectra requires a sufficiently dense grid of tones. Otherwise, a non-negligible portion of the multiband energy resides off the grid, which can lead to recovery errors due to the model mismatch. As discussed below, a dense grid of tones translates to high computational loads in the digital domain.

The MWC is less sensitive to model mismatches in comparison. Since only inequalities are used in (3.25), the number of branches $p$ and aliasing rate $f_p$ can be chosen with some safeguards with respect to the specified number of bands $N$ and individual widths $B$. Thus, the system can handle inputs with more than $N$ bands and widths larger than $B$, up to the safeguards that were set. The band positions are not restricted to any specific displacement with respect to the spectrum slices; a single band can split between slices, as depicted in Fig. 3.6. Nonetheless, both the PNS [19] and MWC [20] approaches require specifying the multiband spectra by a pair of maximal quantities $(N, B)$. This modeling can be inefficient (in terms of resulting sampling rate) when the individual band widths are significantly different from each other. For example, a multiband model with $N_1$ bands of lengths $B_1 = k_1 b$ and $N_2$ bands of lengths $B_2 = k_2 b$ is described by a pair $(N_1 + N_2, \max(B_1, B_2))$, with spectral occupation potentially larger than actually used. A more flexible modeling in this scenario would assume only the total actual bandwidth being occupied, i.e., $N_1 B_1 + N_2 B_2$. This issue can partially be addressed at the expense of hardware size by designing the system (PNS/MWC) to accommodate $N_1 k_1 + N_2 k_2$ bands of lengths $b$.

### 3.8.3 Hardware Complexity

We next compare the hardware complexity of the RD/MWC systems. In both approaches, the acquisition stage is mapped into an underdetermined CS system: Fig. 3.21 leads to a standard sparse recovery problem (3.46) in the RD system, while in the MWC approach, Fig. 3.10 results in an IMV problem (3.12). A crucial point is that the hardware needs to be sufficiently accurate for that mapping to hold, since this is the basis for reconstruction. While the RD and MWC sampling stages seem similar, they rely on different analog properties of the hardware to ensure accurate mapping to CS, which in turn imply different design complexities.

To better understand this issue, we examine Fig. 3.23. The figure depicts the Nyquist-equivalent of each method, which is the system that samples the input at its Nyquist rate and then computes the relevant sub-Nyquist samples by applying the sensing matrix digitally. The RD-equivalent integrates and dumps the input at rate $W$, and then applies $\mathbf{\Phi}$ on $Q$ serial measurements, $\mathbf{x} = [x[1], \cdots, x[Q]]^T$. To coincide with the sub-Nyquist samples of Fig. 3.21, $\mathbf{\Phi} = \mathbf{HD}$ is used, where $\mathbf{D}$ is diagonal with $\pm 1$ entries, according to the values $p_c(t)$ takes on $t = n/W$, and $\mathbf{H}$ sums over $W/R$ entries [30]. The MWC-equivalent has $M$ channels, with the $\ell$th channel demodulating the relevant spectrum slice to the origin and sampling at rate $1/T$, which results in $d_\ell[n]$. The sensing matrix $\mathbf{A}$ is applied on $\mathbf{d}[n]$. While sampling according to the equivalent systems of Fig. 3.23 is a clear waste of resources, it enables us to view the internal mechanism of each strategy. Note that the reconstruction algorithms remain the same; it does not matter whether the samples were actually obtained at a sub-Nyquist rate, according to Figs. 3.21 or 3.10, or if they were computed after sampling according to Fig. 3.23.

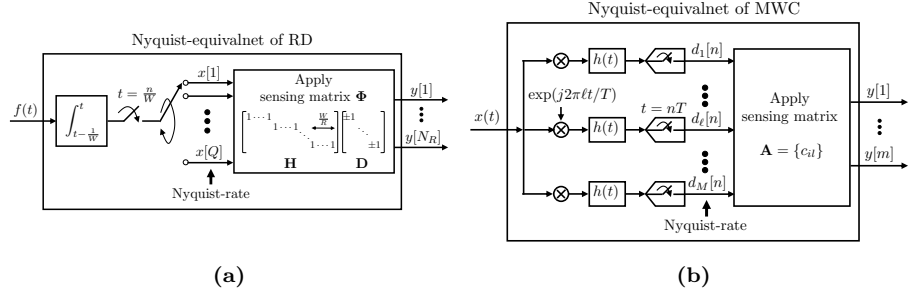

**Figure 3.23:** The Nyquist-equivalents of the RD (a) and MWC (b) sample the input at its Nyquist rate and apply the sensing matrix digitally (taken from [14]).

**Analog compression.** In the RD approach, time-domain properties of the hardware dictate the necessary accuracy. For example, the impulse-response of the integrator needs to be a square waveform with a width of $1/R$ seconds, so that $\mathbf{H}$ has exactly $W/R$ consecutive 1's in each row. For a diagonal $\mathbf{D}$, the sign alternations of $p_c(t)$ need to be sharply aligned on $1/W$ time intervals. If either of these properties is nonideal, then the mapping to CS becomes nonlinear and signal dependent. Precisely, (3.46) becomes [30]

$$\mathbf{y} = \mathbf{H}(\mathbf{x})\mathbf{D}(\mathbf{x})\mathbf{x}. \tag{3.50}$$

A noninteger ratio $W/R$ affects both $\mathbf{H}$ and $\mathbf{D}$ [30]. Since $f(t)$ is unknown, $\mathbf{x}$, $\mathbf{H}(\mathbf{x})$ and $\mathbf{D}(\mathbf{x})$ are also unknown. It is suggested in [30] to train the system on example signals, so as to approximate a linear system. Note that if (3.48) is not satisfied, then the DFT expansion also becomes nonlinear and signal-dependent $\mathbf{x} = \mathbf{F}(\Delta)\mathbf{s}$. The *form factor* of the RD is therefore the time-domain accuracy that can be achieved in practice.

The MWC requires periodicity of the waveforms $p_i(t)$ and lowpass response for $h(t)$, which are both frequency-domain properties. The sensing matrix $\mathbf{A}$ is constant as long as $p_i(t)$ are periodic, regardless of the time-domain appearance of these waveforms. Therefore, nonideal time-domain properties of $p_i(t)$ have no effect on the MWC. The consequence is that stability in the frequency domain dictates the form factor of the MWC. For example, 2 GHz periodic functions were demonstrated in a circuit prototype of the MWC [22]. More broadly, circuit publications report the design of high-speed sequence generators up to 23 and even 80 GHz speeds [85, 86], where stable frequency properties are verified experimentally. Accurate time-domain appearance is not considered a design factor in [85,86], and is in fact not maintained in practice as shown in [22,85,86]. For example, Fig. 3.24 demonstrates frequency stability vs. inaccurate time-domain appearance [22].

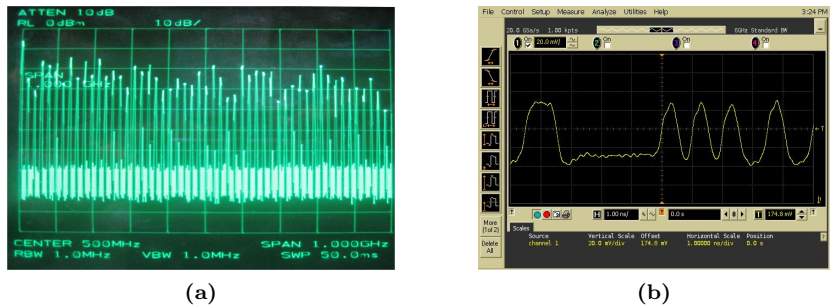

(a)     (b)

**Figure 3.24:** The spectrum (a) and the time-domain appearance (b) of a 2 GHz sign-alternating periodic waveform (taken from [22]).

The MWC scheme requires an ideal lowpass filter $h(t)$ with rectangular frequency response, which is difficult to implement due to its sharp edges. This problem appears as well in Nyquist sampling, where it is addressed by alternative sampling kernels with smoother edges at the expense of oversampling. Similar edge-free filters $h(t)$ can be used in the MWC system with slight oversampling [73]. Ripples in the passband and non-smooth transitions in the frequency response can be compensated for digitally using the algorithm in [59].

**Sampling rate.** In theory, both the RD and MWC approach the minimal rate for their model. The RD system, however, requires in addition an integer ratio $W/R$; see (3.45) and (3.48). In general, a substantial rate increase may be needed to meet this requirement. The MWC does not limit the rate granularity; See a numerical comparison in the next subsection.

**Continuous reconstruction.** The RD synthesizes $\hat{f}(t)$ using (3.43). Realizing (3.43) in hardware can be excessive, since it requires $K$ oscillators, one per each active tone. Computing (3.43) digitally needs a processing rate of $W$, and then a DAC device at the same rate. Thus, the synthesis complexity scales with the Nyquist rate. The MWC reconstructs $\hat{x}(t)$ using commercial DAC devices, running at the low rate $f_s = 1/T$. It needs $N$ branches. Wideband continuous inputs require prohibitively large $K, W$ to be adequately represented on a discrete grid of tones. In contrast, despite the infinitely many frequencies that comprise a multiband input, $N$ is typically small. We note however that the MWC may incurr difficulties in reconstructing contents around the frequencies $(\ell + 0.5)f_p$, $-L \leq \ell \leq L$, since these are irregular points of transitions between spectrum slices. Reconstruction accuracy of these irregular points depends on the cutoff curvature of $h(t)$ and relative amplitudes of consecutive $c_{i\ell}$. Reconstruction of an input consisting of pure tones at these specific frequencies may be imperfect. In practice, the bands encode information signals, which can be reliably decoded, even when signal energy is located around the frequencies $(l + 0.5)fp$. As discussed in Section 3.5.6, when the bands contain digital transmissions and the SNR is sufficiently high, algorithm Back-DSP [14] enables recovery of the underlying information bits, and in turn allows DSP at a low rate, even when a band energy is split between adjacent slices. This algorithm also allows reconstructing $x(t)$ with only $N$ DAC devices instead of $2N$ that are required for arbitrary multiband reconstruction. Table 3.1 summarizes the model and hardware comparison.

Table 3.1: Model and Hardware Comparison

|  | **RD** (multitone) | **MWC** (multiband) |
|---|---|---|
| Model parameters | $K, Q, \Delta$ | $N, B, f_{\max}$ |
| System parameters | $R, W, N_R$ | $m, 1/T$ |
| Setup | (3.45) | (3.25) |
|  | Sensitive, eq. (3.48), Fig. 3.22 | Robust |
| Form factor | time-domain appearance | frequency-domain stability |
| Requirements | accurate $1/R$ integration | periodic $p_i(t)$ |
|  | sharp alternations $p_c(t)$ |  |
| ADC topology | integrate-and-dump | commercial |
| Rate | gap due to (3.45) | approach minimal |
| DAC | 1 device at rate $W$ | $N$ devices at rate $f_s$ |

### 3.8.4 Computational Loads

In this subsection, we compare computational loads when treating multiband signals, either using the MWC system or in the RD framewrod by discretizing the continuous frequency axis to a grid of $Q = f_{\text{NYQ}}$ tones, out of which only $K = NB$ are active [30]. We emphasize that the RD system was designed for multitone inputs, though for the study of computational loads we examine the RD on multiband inputs by considering a comparable grid of tones of the same Nyquist bandwidth. Table 3.2 compares between the RD and MWC for an input with 10 GHz Nyquist rate and 300 MHz spectral occupancy. For the RD we consider two discretization configurations, $\Delta = 1$ Hz and $\Delta = 100$ Hz. The table reveals high computational loads that stem from the dense discretization that is required to represent an analog multiband input. We also included the sampling rate and DAC speeds to complement the previous section. The notation in the table is self-explanatory, though a few aspects are emphasized below.

The sensing matrix $\mathbf{\Phi} = \mathbf{HD}$ of the RD has dimensions

$$\mathbf{\Phi} : R \times W \propto K \times Q \quad \text{(huge)}. \tag{3.51}$$

The dimension scales with the Nyquist rate; already for $Q = 1$ MHz Nyquist-rate input, there are 1 million unknowns in (3.46). The sensing matrix $\mathbf{A}$ of the MWC has dimensions

$$\mathbf{A} : m \times M \propto N \times \frac{f_{\text{NYQ}}}{B} \quad \text{(small)}. \tag{3.52}$$

For the comparable spectral occupancy we consider, $\mathbf{\Phi}$ has dimensions that are 6 to 8 orders of magnitude higher, in both the row and column dimensions, than the MWC sensing matrix $\mathbf{A}$. The size of the sensing matrix is a prominent factor since it affects many digital complexities: the delay and memory length that are associated with collecting the measurements, the number of multiplications when applying the sensing matrix on a vector and the storage requirement of the matrix. See the table for a numerical comparison of these factors.

**Table 3.2:** Discretization Impact on Computational Loads

|  |  | RD |  |  | MWC |  |
|---|---|---|---|---|---|---|
|  | Discretization spacing | $\Delta = 1$ Hz | $\Delta = 100$ Hz |  |  |  |
| Model | $K$ tones | $300 \cdot 10^6$ | $3 \cdot 10^6$ | $N$ bands |  | 6 |
|  | out of $Q$ tones | $10 \cdot 10^9$ | $10 \cdot 10^7$ | width $B$ |  | 50 MHz |
| Sampling setup | alternation speed $W$ | 10 GHz | 10 GHz | $m$ channels[S] |  | 35 |
|  |  |  |  | $M$ Fourier coefficients |  | 195 |
|  | rate $R$, eq. (3.47), theory | 2.9 GHz | 2.9 GHz | $f_s$ per channel |  | 51 MHz |
|  | eq. (3.45), practice | 5 GHz | 5 GHz | total rate |  | 1.8 GHz |
| Underdetermined system |  | (3.46): $\mathbf{y} = \mathbf{HDFs}$, $\|\mathbf{s}\|_0 \leq K$ |  | (3.19): $\mathbf{V} = \mathbf{AU}$, $\|\mathbf{U}\|_0 \leq 2N$ |  |  |
| Preparation |  |  |  |  |  |  |
| Collect samples | Num. of samples $N_R$ | $5 \cdot 10^9$ | $5 \cdot 10^7$ | $2N$ snapshots of $\mathbf{y}[n]$ |  | $12 \cdot 35 = 420$ |
| Delay | $N_R/R$ | 1 sec | 10msec | $2N/f_s$ |  | 235nsec |
| Complexity |  |  |  |  |  |  |
| Matrix dimensions | $\mathbf{\Phi} = \mathbf{HDF} = N_R \times Q$ | $5 \cdot 10^9 \times 10^{10}$ | $5 \cdot 10^7 \times 10^8$ | $\mathbf{A} = m \times M$ |  | $35 \times 195$ |
| Apply matrix[♯] | $\mathcal{O}(W \log W)$ |  |  | $\mathcal{O}(mM)$ |  |  |
| Storage[♯] | $\mathcal{O}(W)$ |  |  | $\mathcal{O}(mM)$ |  |  |
| <u>Realtime</u> (fixed support) |  | $\mathbf{s}_\Omega = (\mathbf{\Phi F})^\dagger_\Omega \mathbf{y}$ |  | $\mathbf{d}_\lambda[n] = \mathbf{A}^\dagger_\lambda \mathbf{y}[n]$ |  |  |
| Memory length | $N_R$ | $5 \cdot 10^9$ | $5 \cdot 10^7$ | 1 snapshot of $\mathbf{y}[n]$ |  | 35 |
| Delay | $N_R/R$ | 1 sec | 10msec | $1/f_s$ |  | 19.5nsec |
| Mult.-ops. (per window) | $KN_R$ | $1.5 \cdot 10^{18}$ | $1.5 \cdot 10^{14}$ | $2Nm$ |  | 420 |
| (100 MHz cycle) | $KN_R/((N_R/R) \cdot 100\text{M})$ | $1.5 \cdot 10^{10}$ | $1.5 \cdot 10^6$ | $2Nmf_s/100\text{M}$ |  | 214 |
| Reconstruction |  | 1 DAC at rate $W = 10$ GHz |  | $N = 6$ DACs at individual rates $f_s = 51$ MHz |  |  |
| Technology barrier (estimated) |  | CS algorithms ($\sim$10 MHz) |  | Waveform generator ($\sim$23 GHz) |  |  |

[S] with $q = 1$; in practice, hardware size is collapsed with $q > 1$ [22].     ♯ for the RD, taking into account the structure $\mathbf{HDF}$.

We also compare the reconstruction complexity, in the more simple scenario that the support is fixed. In this setting, the recovery is merely a matrix-vector multiplication with the relevant pseudo-inverse. As before, the size of $\mathbf{\Phi}$ results in long delay and huge memory length for collecting the samples. The number of scalar multiplications (Mult.-ops.) for applying the pseudo-inverse reveals again orders of magnitude differences. We expressed the Mult.-ops. per block of samples, and in addition scaled them to operations per clock cycle of a 100 MHz DSP processor.

We conclude the table with our estimation of the technology barrier of each approach. Computational loads and memory requirements in the digital domain are the bottleneck of the RD approach. Therefore the size of CS problems that can be solved with available processors limits the recovery. We estimate that $W \approx 1$ MHz may be already quite demanding using convex solvers, whereas $W \approx 10$ MHz is probably the barrier using greedy methods[3]. The MWC is limited by the technology for generating the periodic waveforms $p_i(t)$, which depends on the specific choice of waveform. The estimated barrier of 23 GHz refers to implementation of the periodic waveforms according to [85, 86], though realizing a full MWC system at these high rates can be a challenging task. Our barrier estimates are roughly consistent with the hardware publications of these system: [88, 89] report the implementation of (single, parallel) RD for Nyquist-rate $W = 800$ kHz. An MWC prototype demonstrates faithful reconstruction of $f_{\text{NYQ}} = 2$ GHz wideband inputs [22].

---

[3] A bank of RD channels was studied in [87], the parallel system duplicates the analog issues and its computational complexity is not improved by much.

### 3.8.5   Analog vs. Discrete CS Radar

The question of whether finite modeling can be used to treat general analog scenarios was also studied in [29] in the context of radar imaging. Here, rate reduction can be translated to increased resolution and decreased time-bandwidth product of the radar system.

An intercepted radar signal $x(t)$ has the form

$$x(t) = \sum_{k=1}^{K} \alpha_k h(t - t_k) e^{j 2\pi \nu_k t} \qquad (3.53)$$

with each triplet $(t_k, \nu_k, \alpha_k)$ corresponding to an echo of the radar waveform $h(t)$ from a distinct target [90]. Equation (3.53) represents an infinite union, parameterized by $\lambda = (t_k, \nu_k)$, of $K$-dimensional subspaces $\mathcal{A}_\lambda$ which capture the amplitudes $\alpha_k$ within the chosen subspace. The UoS approach was taken in [29], where reconstruction is obtained by the general scheme for time delay recovery of [26], with subspace estimation that uses standard spectral estimation tools [70]. A finite modeling approach to radar assumes that the delays $t_k$ and frequencies $\nu_k$ lie on a grid, effectively quantizing the delay–Doppler space $(t, \nu)$ [32, 33, 91]. CS algorithms are then used for reconstruction of the targets scene.

An example of identification of nine targets (in a noiseless setting) is illustrated in Fig. 3.25 for three approaches: the union-based approach [29] with a simple lowpass acquisition, classic matched filtering and quantized-CS recovery. The discretization approach causes energy leakage in the quantized space into adjacent grid points. As the figure shows, union modeling is superior with respect to both alternative approaches. Identification results in the presence of noise appear in [29] and affirm imaging performance that degrades gracefully as noise levels increase, as long as the noise is not too large. These results affirm that UoS modeling not only offers a reduced-rate sampling method, but allows to increase the resolution in target identification, as long as the noise is not too high. At high noise levels, match-filtering is superior. We refer to [76] for rigorous analysis of noise effects in general FRI models.

A property of great interest in radar applications is the time-bandwidth $\mathcal{WT}$ product of the system, where $\mathcal{W}$ refers to the bandwidth of the transmitted pulse $h(t)$ and $\mathcal{T}$ indicates the round-trip time between radar station and targets. Ulti-mately we would like to minimize both quantities, since $\mathcal{W}$ impacts antenna size and sampling rate, while $\mathcal{T}$ poses a physical limitation on latency, namely the time it takes to identify targets. Uncertainty principles, though, imply that we cannot minimize both $\mathcal{W}$ and $\mathcal{T}$ simultaneously. The analog CS radar approach results in minimal time-bandwidth product, much lower than that obtained using standard matched-filtering techniques; see [29] for a precise comparison. Practical aspects of sparsity-based radar imaging, such as improved decoding time of target identification from compressive measurements as well as efficient matrix structures for radar sensing, are studied in [92].

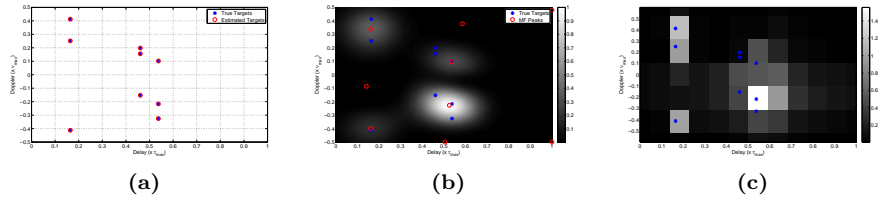

**Figure 3.25:** Recovery of the Doppler-delay plane using (a) a union of subspaces approach, (b) a standard matched filter, and (c) a discretized delay-Doppler plane (taken from [29]).

## 3.9  Discussion

Table 3.3 summarizes the various applications we surveyed, suggesting that Xampling is broad enough to capture a multitude of engineering solutions, under the same logical flow of operations. We conclude with a discussion on the properties and insights into analog sensing highlighted throughout this chapter.

### 3.9.1  Extending CS to Analog Signals

The influential works by Donoho [3] and Candès et al. [4] coined the CS terminology, in which the goal is to reduce the sampling rate below Nyquist. These pioneering works established CS via a study of underdetermined systems, where the sensing matrix abstractly replaces the role of the sampling operator, and the ambient dimensions represent the high Nyquist rate. In practice, however, the study of underdetermined systems does not hint at the actual sub-Nyquist sampling of analog signals. One cannot apply a sensing matrix on a set of Nyquist rate samples, as performed in the conceptual systems in Fig. 3.23, since that would contradict the whole idea of reducing the sampling rate. The previous sections demonstrate how extensions of CS to continuous signals can be significantly different in many practical aspects. Based on the insights gained, we draw several operative conclusions in Table 3.4 regarding the choice of analog com-pression $P$ in continuous-time CS systems. The first point follows immediately from Fig. 3.22 and basically implies that model and sampler parameters should not be tightly related, implicitly or explicitly. We elaborate below on the other two suggestions.

Table 3.3: Applications of union of subspaces

| Application | Signal model | Cardinality union | Cardinality subspaces | Analog compression | Subspace detection |
|---|---|---|---|---|---|
| Sparse-SI [16] | see (3.11) | finite | $\infty$ | filter-bank, Fig. 3.4 | CTF |
| PNS [19] | multiband, Fig. 3.6 | finite | $\infty$ | time shifts | CTF [44] |
| MWC [20] | multiband, Fig. 3.6 | finite | $\infty$ | periodic mixing + lowpass | CTF [44] |
| RD [30] | $f(t) = \sum_\omega a_\omega e^{j2\pi\omega t}$, $\omega \in$ discrete grid $\Omega$ | finite | finite | sign flipping + integrate-dump | CS |
| FRI | $x(t) = \sum_{\ell=1}^{L} d_\ell\, g(t-t_\ell)$ | | | | |
|   periodic [24, 93] | $x(t) = x(t+T)$ | $\infty$ | finite | lowpass | annihilating filter [24, 93] |
|   finite [25] | $0 \leq t \leq T$ | $\infty$ | finite | splines | moments factoring [25] |
|   periodic/finite [27, 28] | either of the above | $\infty$ | finite | SoS filtering | annihilating filter |
| Sequences of innovation [26, 28] | see (3.40) | $\infty$ | $\infty$ | lowpass, or periodic mixing + integrate-dump | MUSIC [5] or ESPRIT [6] |
| NYFR [94] | multiband | finite | $\infty$ | jittered undersampling (nonlinear) | n/a |

Table 3.4: Suggested Guidelines for Extending CS to Analog Signals

| | |
|---|---|
| #1 | set system parameters with safeguards to accommodate possible model mismatches |
| #2 | incorporate design constraints on $P$ that suit the technology generating the source signals |
| #3 | balance between nonlinear (subspace detection) and linear (interpolation) reconstruction complexities |

Input signals are eventually generated by some source, which has its own accuracy specifications. Therefore, if designing $P$ imposes constraints on the hardware that are not stricter than those required to generate the input signal, then there are no essential limitations on the input range. We support this conclusion by several examples. The MWC requires accuracy that is achieved with RF technology, which also defines the possible range of multiband transmissions. The same principle of shifting spectral slices to the origin with different weights can be achieved by PNS [19]. This strategy, however, can result in a narrower input range that can be treated, since current RF technology can generate source signals at frequencies that exceed front-end bandwidths of existing ADC devices [20]. Multiband inputs generated by optical sources, however, may require a different compression stage $P$ than that of the RF-based MWC system.

Along the same line, time-domain accuracy constraints may limit the range of multitone inputs that can be treated in the RD approach, if these signals are generated by RF sources. On the other hand, consider a model of piecewise constant inputs, with knots at the integers and only $K$ nonidentically-zero pieces out of $Q$. Sampling these signals with the RD system would map to (3.46), but with an identity basis instead of the DFT matrix $\mathbf{F}$. In this setting, the time-domain accuracy required to ensure that the mapping to (3.46) holds is within the tolerances of the input source.

Moving on to our third suggestion, we attempt to reason the computational loads encountered in Table 3.2. Over 1 second, both approaches reconstruct their inputs from a comparable set of numbers; $K = 300 \cdot 10^6$ tone coefficients or $2Nf_s = 612 \cdot 10^6$ amplitudes of active sequences $d_\ell[n]$. The difference is, however, that the RD recovers all these unknowns by a single execution of a nonlinear CS algorithm on the system (3.46), which has large dimensions. In contrast, the MWC splits the recovery task to a small-size nonlinear part (*i.e.,* CTF) and real-time linear interpolation. This distinction can be traced back to model assumptions. The nonlinear part of a multitone model, namely the number of subspaces $|\Lambda| = \binom{Q}{K}$, is exponentially larger than $\binom{M}{2N}$ which specifies a multiband union of the same Nyquist bandwidth. Clearly, a prerequisite for balancing computation loads is an input model with as many unknowns as possible in its linear part (subspaces $\mathcal{A}_\lambda$), so as to decrease the nonlinear cardinality $|\Lambda|$ of the union. The important point is that in order to benefit from such modeling, $P$ must be properly designed to incorporate this structure and reduce computational loads.

For example, consider a block-sparse multitone model with $K$ out of $Q$ tones, such that the active tones are clustered in $K/d$ blocks of length $d$. A plain RD system which does not incorporate this block structure would still result in a large $R \times W$ sensing matrix with its associated digital complexities. Block-sparse recovery algorithms, *e.g.,* [42], can be used to partially decrease the complexity, but the bottleneck remains the fact that the hardware compression is mapped to a large sensing matrix[4]. A potential analog compression for this block-sparse model can be an MWC system designed for $N = K/d$ and $B = d\Delta$ specifications.

Our conclusions here stem from the study of the RD and MWC systems, and are therefore mainly relevant for choosing $P$ in Xampling systems that map their hardware to underdetermined systems and incorporate CS algorithms for recovery. Nonetheless, our suggestions above do not necessitate such a relation to CS, and may hold more generally with regard to other compression techniques.

Finally, we point out the Nyquist-folding receiver (NYFR) of [94] which suggests an interesting alternative route towards sub-Nyquist sampling. This method introduces a deliberate jitter in an undersampling grid, which results in induced phase modulations at baseband such that the modulation magnitudes depend on the unknown carrier positions. This strategy is exceptional as it relies on a nonlinear acquisition effect, which departs from the linear $P$ that has been utilized in all previous applications. In principle, to enable recovery, one would need to infer the magnitudes of the phase modulations. A reconstruction algorithm was not reported yet for this class of sampling, which is why we do not elaborate further on this method. Nonetheless, the innovative idea of using nonlinear compression $P$ opens a wide range of possibilities to explore.

---

[4] Note that simply modifying the chipping and integrate-dumping intervals, in the existing scheme of Fig. 3.21, to $d$ times larger results in a sensing matrix smaller by the same factor, though (3.46) in this setting would force reconstructing each block of tones by a single tone, presumably corresponding to a model of $K/d$ active tones out of $Q/d$ at spacing $d\Delta$.

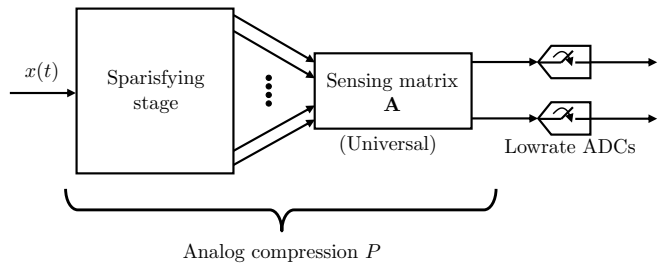

**Figure 3.26:** Analog compression operator $P$ in X-ADC architecture consists of a sparisfying stage and sensing matrix, which are combined into one efficient analog preprocessing stage.

### 3.9.2   Is CS a Universal Sampling Scheme ?

The discussion on extending CS to analog signals draws an interesting connection to the notion of CS universality. In the discrete setup of sensing, the measurement model is $\mathbf{y} = \mathbf{\Phi x}$ and the signal is sparse in some given transform basis $\mathbf{x} = \mathbf{\Psi s}$. The concept of CS universality refers to the attractive property of sensing with $\mathbf{\Phi}$ without knowledge of $\mathbf{\Psi}$, so that $\mathbf{\Psi}$ enters only in the reconstruction algorithm. This notion is further emphasized with the default choice of the identity basis $\mathbf{\Psi} = \mathbf{I}$ in many CS publications, which is justified by no loss of generality, since $\mathbf{\Psi}$ is conceptually absorbed into the sensing matrix $\mathbf{\Phi}$.

In contrast, in many analog CS systems, the hardware design benefits from incorporating knowledge on the sparsity basis of the input. Refer to the Nyquist-equivalent system of the MWC in Fig. 3.23(b), for example. The input $x(t)$ is conceptually first preprocessed into a set of high-rate streams of measurements $\mathbf{d}[n]$, and then a sensing matrix $\mathbf{A} = \{c_{i\ell}\}$ is applied to reduce the rate. In PNS [20], the same set of streams $\mathbf{d}[n]$ is sensed by the partial DFT matrix (3.23), which depends on the time shifts $c_i$ of the PNS sequences. This sensing structure also appears in Theorem 3.1, where the term $\mathbf{G}^{-*}(e^{j\omega T})\mathbf{s}(\omega)$ in (3.13) first generates $\mathbf{d}[n]$, and then a sensing matrix $\mathbf{A}$ is applied. In all these scenarios, the intermediate sequences $\mathbf{d}[n]$ are sparse for all $n$, so that the sensing hardware effectively incorporates knowledge on the (continuous) sparsity basis of the input.

Figure 3.26 generalizes this point. The analog compression stage $P$ in Xampling systems can be thought of a two stages sampling system. First, a sparsifying stage which generates a set of high-rate streams of measurements, out of which only a few are nonidentically zero. Second, a sensing matrix is applied, where in principle, any sensing matrix can be used in that stage. In practice, however, the trick is to choose a sensing matrix which can be combined with the sparsifying part into a single hardware mechanism, so that the system does not actually go through Nyquist-rate sampling. This combination is achieved by periodic mixing in the MWC system, time-delays in the case of PNS, and the filters $w_\ell(t)$ in the sparse-SI framework of Theorem 3.1. We can therefore suggest a slightly different interpretation of the universality concept for analog CS systems, which is the flexibility to choose any sensing matrix $\mathbf{A}$ in the second stage of $P$, provided that it can be efficiently combined with the given sparsifying stage.

### 3.9.3 Concluding Remarks

Starting from the work in [15], union of subspaces models appear at the frontier of research on sampling methods. The ultimate goal is to build a complete sampling theory for UoS models of the general form (3.3) and then derive specific sampling solutions for applications of interest. Although several promising advances have already been made [15,16,26,40,43], this esteemed goal is yet to be accomplished.

In this chapter we described a line of works which extend CS ideas to the analog domain based on UoS modeling. The Xampling framework of [14] unifies the treatment of several classes of UoS signals, by leveraging insights and pragmatic considerations into the generic architecture of Fig. 3.2.

Our hope is that the template scheme of Fig. 3.2 can serve as a substrate for developing future sampling strategies for UoS models, and inspire future developments that will eventually lead to a complete generalized sampling theory in unions of subspaces.

## Acknowledgements

The authors would like to thank Kfir Gedalyahu and Ronen Tur for their collaboration on many topics related to this review and for authorizing the use of their figures, and Waheed Bajwa for many useful comments.